\documentclass{article}

\usepackage{arxiv}

\usepackage[utf8]{inputenc} 
\usepackage[T1]{fontenc}    
\usepackage{hyperref}       
\usepackage{url}            
\usepackage{booktabs}       
\usepackage{amsfonts}       
\usepackage{nicefrac}       
\usepackage{microtype}      
\usepackage{lipsum}
\usepackage{comment}

\usepackage{amsmath}		
\usepackage{placeins}
\usepackage[caption=false]{subfig}
\usepackage[noend]{algpseudocode}

\usepackage{enumitem}
\usepackage{graphicx}
\usepackage{makecell}
\usepackage{tabularx}
\usepackage{caption}    

\usepackage[ruled]{algorithm2e} 
\DeclareMathOperator*{\argmin}{argmin}
\DeclareMathOperator*{\argmax}{argmax}

\title{Dynamic Parameter Estimation of Brain Mechanisms}

\author{
  Po-Ya Hsu\thanks{poyahsu.com} \\
  Department of Computer Science\\
  University of California, San Diego\\
  La Jolla, CA 92093 \\
  \texttt{p8hsu@eng.ucsd.edu}\\
}

\begin{document}
\maketitle

\begin{abstract}
Demystifying effective connectivity among neuronal populations has become the trend to understand the brain mechanisms of Parkinson's disease, schizophrenia, mild traumatic brain injury, and many other unlisted neurological diseases.
Dynamic modeling is a state-of-the-art approach to explore various connectivities among neuronal populations corresponding to different electrophysiological responses.
Through estimating the parameters in the dynamic models, including the strengths and propagation delays of the electrophysiological signals, the discovery of the underlying connectivities can lead to the elucidation of functional brain mechanisms.
In this report, we survey six dynamic models that describe the intrinsic function of a single neuronal/subneuronal population and three effective network estimation methods that can trace the connections among the neuronal/subneuronal populations. The six dynamic models are event related potential, local field potential, conductance-based neural mass model, mean field model, neural field model, and canonical micro-circuits; the three effective network estimation approaches are dynamic causal modeling, structural causal model, and vector autoregression.
Subsequently, we discuss dynamic parameter estimation methods including variational Bayesian, particle filtering, Metropolis-Hastings algorithm, Gauss-Newton algorithm, collocation method, and constrained optimization.
We summarize the merits and drawbacks of each model, network estimation approach, and parameter estimation method.
In addition, we demonstrate an exemplary effective network estimation problem statement.
Last, we identify possible future work and challenges to develop an elevated package.
\end{abstract}


\section{Introduction}

Demystifying effective connectivity among neuronal populations has become the trend to understand the dynamic brain mechanisms such as Parkinson's disease and mismatch negativity, and some sophisticated techniques have been proposed to reconstruct the underlying neuronal networks \cite{razi2016connected,van2013wu,insel2013nih,bressler2010large}.
In fact, studies of connectivities inside brains has a long history, and it was not until the past decades that effective connectivity was recognized as the concept behind the functioning brain.

\subsection{Terminology}
A few vocabulary words or collocations can be difficult to interpret due to their usage in interdisciplinary fields. Therefore, this subsection serves as the clarification for nuisance removal.
\begin{itemize}
    \item Neural Network \& Neuronal Network: Neural network is the network of nervous system, however, it also represents artificial neural network in computer science. In this sense, neuronal network is chosen to represent biological neural network for ease of reading in this report.
    \item Effective Connectivity \& Functional Connectivity: An intuitive way to separate the two terms is to verify whether the parameters in the dynamic models are being used to interpret the observations or not. If yes, the model belongs to effective connectivity.   
    \item Functional Integration \& Functional Segregation: Functional integration allows different cortical areas in the brain to interchangeably pop up and disappear, whereas in functional segregation, the activation of cortical areas is preassigned.
    \item Cortical Column: Cortical column is the hierarchical structure found in the cortex that represents the basic unit of functioning. More details are provided in Section \ref{sec:formulation}.
    \item Neuronal Population: In this report, neuronal population is a population of sufficiently large number of neuronal cells of the same type.
    \item Mismatch negativity: Mismatch negativity is a negative change in the auditory evoked potential secondary to the unpredictable deviated auditory stimuli. For example, a person gets used to hearing 1k Hertz stimuli. Following the regular 1k Hertz stimuli, a 1.1k Hertz stimulus is sent to the person, and the change found in EEG/MEG (differed from the response of 1k Hertz) is what we called mismatch negativity.  
    \item Alpha Activity: A quasi-periodic electrical signal around 8-12 Hertz is observed in waking people with eyes closed, which has been believed to originate in occipital lobe.
\end{itemize}

\subsection{Related Work}

Before the $20^{th}$ century, neuroscientists believed that functional segregation, or functional localization, gave explanations to brain mechanisms \cite{razi2016connected,sharkey2009connectionism}. Functional segregation is an idea that each cortical region in the brain is specialized for certain tasks only \cite{tononi1994measure}. By analogy, in computers, each program has already been preassigned to specific tasks. Functional localization, which is different from functional segregation, takes on a more general aspect. Functional localization implies that a task can be localized in the corresponding cortical areas, which means the identical cortical area can be paired with multiple tasks. However, with the incremental discovery in adaptability, flexibility, and environmentally dependent modality of the brain \cite{cole2013multi}, the concept of brain mechanisms transformed into the internal connection of the brain networks. 

In the 1940s, the word \textit{connectionism} was coined by Donald Hebb to describe the causal relation between the memory and the perception to the physical world \cite{sharkey2009connectionism,ketchum1959mind}. Hebb proposed that the synaptic connection between the neurons in a certain path facilitates the perception to a specific pattern. Starting publishing the initial report in 1959, Hubel and Wiesel, recipients of the Nobel Prize in Physiology or Medicine 1981, corroborated the distributed process in the visual cortex of the monkey with over 25 years research conducted \cite{hubel1974uniformity,wiesel1974ordered}. In their experiments, Hubel and Wiesel found how visual perception was produced by well-organized cortical neurons \cite{wurtz2009recounting}. Their contribution not only led to an explosion in the number of studies beyond visual cortex, but also inspired the developments of machine learning and deep learning in computer science \cite{george2009towards,lecun2015deep}.

Currently, based on the inception of brain's adaptibility, flexibility, and modality, neuroscientists accept both functional integration and functional segregation as how the brain works \cite{friston2002functional,macaluso2005multisensory}. Functional integration indicates that brain mechanism is a dynamic self-assembling process. In other words, within a specified time span in an activity, parts of the brain engage and disengage in the activity temporally. For example, in mismatch negativity, the auditory cortex reacts to the stimuli strongly at initial phase, but during later period, the magnitude is suppressed due to the brain's adaptability \cite{garrido2009mismatch}. To resolve connectivities among neuronal populations under the assumption of functional integration, two main approaches are adopted widely by neuroscientists: functional connectivity and effective connectivity.

The dichotomy between functional connectivity and effective connectivity relies on the consideration of dynamic models \cite{friston2011functional}. Functional connectivity is determined by the measures of statistical dependencies such as Pearson correlations and transfer entropy \cite{greicius2003functional,wibral2014transfer}. In contrast to functional connectivity, effective connectivity depends on the parameters of dynamic models that can interpret the observed statistical dependencies (functional connectivity). In \cite{aertsen1991dynamics}, effective connectivity is defined as the \textit{ time-dependent and simplest possible circuit diagram that would replicate the observed temporal relationship between the recorded neurons.} In this regard, effective connectivity infers more information on causal relationships between the neuronal populations than functional connectivity does.



\subsection{Motivation}
In system biology, dynamic modeling often consists of differential equations and observation formulation \cite{villaverde2014reverse}; hence, the whole system can be expressed with differential equations $f$ that express the dynamic transformation of (state) variables $\mathbf{x}$ with parameters $\mathbf{\Theta}$, inputs $\mathbf{u}(t)$ as arguments, and $\mathbf{w}(t)$ as endogenous fluctuations, and observation function $g$ that generates data $\mathbf{y}(t)$ depending on variables $\mathbf{x}$, observation noise $\mathbf{e}(t)$ and parameters $\mathbf{\Theta}$:
\begin{align}
\label{eq:inverse}
\dot{x} &= f(\mathbf{x}(t),\mathbf{u}(t),\mathbf{\Theta})+\mathbf{w}(t), \\
\mathbf{y}(t) &= g(\mathbf{x}(t),\mathbf{\Theta})+\mathbf{e}(t).
\end{align}

The framework for a generalized reverse engineering problem in system biology is suitable for modeling coupled neuronal networks generated by neuronal populations \cite{david2006dynamic,razi2016connected}. Vector $\mathbf{x}$ stands for the currents or potentials of neuronal populations; vector $\mathbf{y}$ is the observed neuroimaging data. Suppose vector $\mathbf{x}$ has $n$ elements, then there are $n$ hidden state variables in vector $\mathbf{x}$. Parameters $\mathbf{\Theta}$ are the parameterized connectivities, which are the targets to be approximated in this report. Inputs $\mathbf{u}(t)$ are often deterministic such as visual cue or auditory stimuli. Function $g$ depends on the instruments and the types of the neuroimaging data, but is not discussed in the current report.

This report aims at a survey of effective connectivity estimation, which corresponds to seeking parameters $\mathbf{\Theta}$ in differential equations $f$.
Effective connection consists of dynamic models that can describe the mechanisms of neuronal populations and network estimation approaches that can fine-tune the parameters of the networks. In Section \ref{sec:NPM}, six dynamic models (local field potential, event related potential, neural mass model, neural field model, mean field model, and canonical micro-circuits) that describe the intrinsic function of a single neuronal/subneuronal population are surveyed.
In Section \ref{sec:ENN}, three effective network estimation methods (dynamic causal modeling, structural causal model, and vector autoregression) that can trace the connections among the neuronal/subneuronal populations are investigated.
In Section \ref{sec:estimation}, we survey dynamic parameter estimation methods including variational Bayesian, particle filtering, Metropolis-Hastings algorithm, Gauss-Newton algorithm, collocation method, and constrained optimization.
For each surveyed dynamic model, network estimation approach, and parameter estimation method, we summarize its advantages and disadvantages and point out the challenges. Following the surveyed contents, we demonstrate an exemplary effective network problem statement in Section \ref{sec:formulation}.
Last, we identify possible future work and make conclusions in Sections \ref{sec:future} and \ref{sec:conclude} respectively.


\section{Neuronal Population Models}
A suite of dynamic models for the intrinsic network in a cortical area have been proposed, and in this report, the focus lays on the models designed for electroencephalography (EEG) and magnetoencephalography (MEG). Both EEG and MEG offer aggregate measure of neuronal activities with 1 millisecond resolution \cite{baillet2001electromagnetic}. According to \cite{moran2013neural}, the models that can quantify the measured neural activities of EEG/MEG include convolution-based neural mass models, conductance-based neural mass models, and field-based neural mass models. These models represent function $f$ in Equation (1), and should be adopted appropriately in distinct circumstances.  

\label{sec:NPM}

\subsection{Convolution-based Neural Mass Model} 
\begin{figure}
  \centering
    \includegraphics[width=1.0\linewidth]{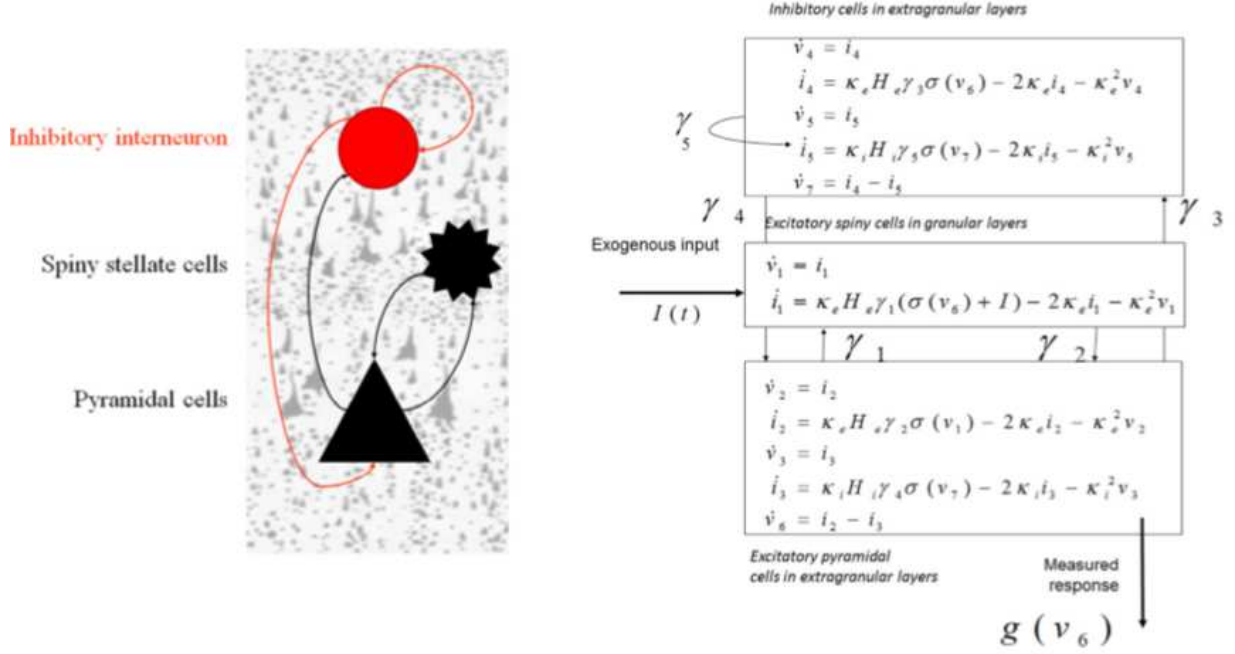}
  \caption{Convolution-based neural mass models, figure source(\cite{moran2013neural})} 
  \label{fig:convNMM}
\end{figure}

\hfill
\\
Convolution-based neural mass models are constructed under the assumption that each cortical column (composed of a certain amount of hierarchical neurons) can be treated as a point of mass \cite{freeman1987simulation,jansen1995electroencephalogram,wendling2000relevance}.
The function of mass-based neural models lies in the sophisticated circuit composed of three subneuronal populations, and Figure \ref{fig:convNMM} shows the differential equations and biological demonstration of convolution-based neural mass models. The cortical column is divided into three subneuronal populations based on experiments in macaque monkeys conducted by Felleman and associate \cite{felleman1991distributed}. The top layer, the supragranular layer (layers 1-3), is consisted of GABAergic inhibitory interneurons; the middle layer, the granular layer 4 (layer 4), is composed of spiny stellate cells; the bottom layer, the infragranular layer (layers 5-6), is formed by pyramidal cells. The interconnection among the three subpopulation cells is also depicted in Figure \ref{fig:convNMM}. The excitatory spiny stellate cells receive inputs and send the signals to pyramidal cells. Subsequently, excitatory pyramidal cells pass the message up to inhibitory interneurons. The inhibitory interneurons then regulate the process with feedback sent to pyramidal cells and sequential relay to spiny stellate cells. The reciprocal process establishes the fundamental message passing in nervous system. 

The differential equations for one neuronal population in convolution-based neural mass models can be written as \cite{moran2013neural}:
\begin{align}
\dot{v}(t) &= i(t), \\
\dot{i}(t) &= \kappa_{e/i} H_{e/i} \gamma S(v_{off}) - 2 \kappa_{e/i} i(t) -\kappa_{e/i}^2 v(t).    
\end{align}
Potential $v$ and current $i$ are elements of vector $x$. Function $S$ is the sigmoid function, $H_e \ , \kappa_e \ , H_i, \ \kappa_i, \ $ are the parameters for membrane potentials, and $\gamma$ is the coefficient for internal connection. $v_{off}$ is the input to the neuronal population. The derivation of differential equations (3) - (4) are based on the neural mass model proposed by Jansen and Rit in 1995 \cite{jansen1995electroencephalogram}. In neural mass model, for each neuronal population, the second order differential equation 
\begin{equation}
\ddot{x}(t)=Aau(t)-2a\dot{x}(t)-a^2x(t)
\end{equation} transforms average density of presynaptic inputs to postsynaptic membrane potential (PSP), where $u(t)$ and $x(t)$ are the presynaptic input and postsynaptic output signals, respectively; $A$, $a$ are the amplitude and decay time respectively. The solution of excitatory neuronal populations is
\begin{equation}
h_e(t) = 
  \begin{cases}
    \frac{H_e}{\tau_e} te^{- \frac{t}{\tau_e} } \, , \, t \geq 0 \\
     0 \, , \, t< 0 .
  \end{cases}
\end{equation}
On the other hand, the solution of inhibitory interneurons is
\begin{equation}
h_i(t) = 
  \begin{cases}
    \frac{H_i}{\tau_i} te^{- \frac{t}{\tau_i} } \, , \, t \geq 0 \\
     0 \, , \, t< 0 
  \end{cases}
\end{equation}
In nervous system, to successfully fire the neuron and complete signal propagation, the net presynaptic inputs must be larger than the threshold. Such threshold-dependent mechanism is integrated with a modified sigmoid function: 
\begin{equation}
S(v) = \frac{1}{1+exp(-rv)}-\frac{1}{2},
\end{equation}
where $r=0.56$.


\paragraph{\textbf{Event Related Potential}} \hfill

Event related potential (ERP) model belongs to convolution-based neural mass model, and ERP model has been utilized in the study of mismatch negativity \cite{garrido2008functional}. ERP is modeled as the dynamic response owing to the exogenous input; ERP has the structure similar to Equations in Figure \ref{fig:convNMM} except the regulation in the layer composed of inhibitory interneurons. In this regard, equations for $v_5,i_5,v_7$ are deducted from the full equations displayed in Figure \ref{fig:convNMM} in ERP model. For ERP's application in mismatch negativity, researchers have tested various hypothesized network structures (on the scale of cortical columns) with ERP model. Competing ERP network models have been compared to learn the mutable neuronal networks in studies of mismatch negativity \cite{boly2011preserved,naatanen2005memory}.

\paragraph{\textbf{Local Field Potential}} \hfill

Local field potential (LFP) model is derived from ERP model, and LFP focuses on capturing the steady-state spectrum (through Fourier Transform) of time series data \cite{friston2012dcm}. Unlike the ERP model, LFP model includes the regulation of inhibitory interneurons ($v_5,i_5,i_7$). LFP model was based on the biophysical proposed by Whittington et al. to simulate the gamma oscillation in the hippocampus \cite{whittington1995synchronized}. Currently, researchers have been making attempts on making inference about synaptic functions by adopting the LFP model \cite{moran2008bayesian}.

\subsection{Conductance-based Neural Mass Model} 
\begin{figure}[!htpb]
  \centering
    \includegraphics[width=1.0\linewidth]{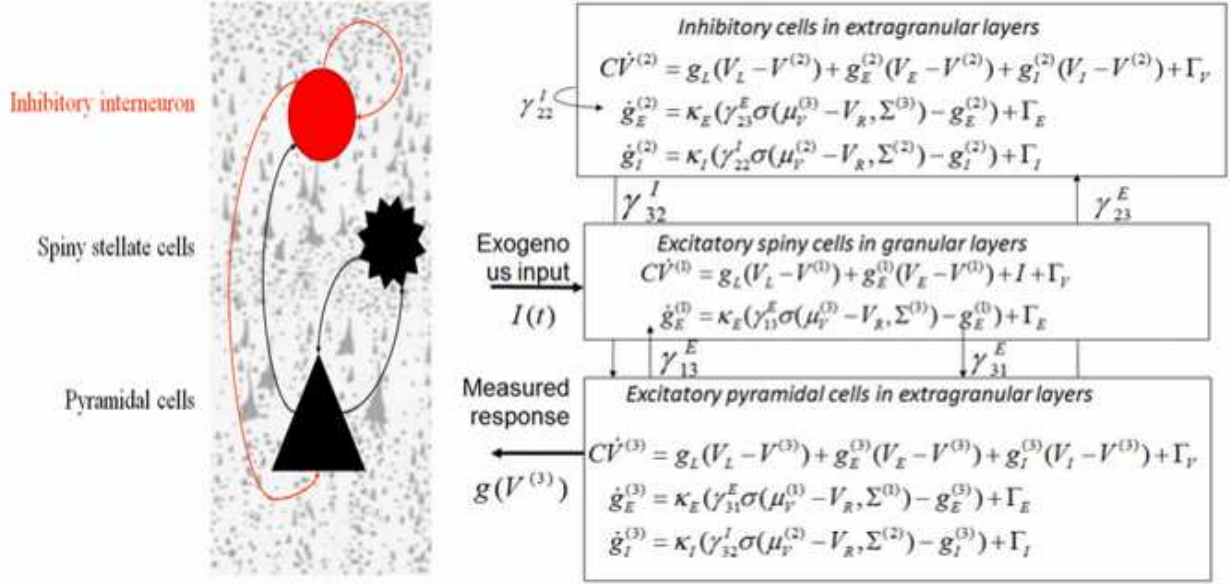}
  \caption{Conductance-based neural mass models, figure source(\cite{moran2013neural})} 
  \label{fig:conductNMM}
\end{figure}

\hfill
\\
Conductance-based neural mass models are constructed on the biophysical models for describing the electrical properties of a single neuron and provide more information than convolution-based neural mass models. Figure \ref{fig:conductNMM} shows the differential equation set and the biological framework of the conductance-based neural mass model built on Morris-Lecar model. 

Hodgkin-Huxley membrane model, Morris-Lecar model, and Fitzhugh-Nagumo model are three typical biophysical models to simulate the initiation and propagation of action potentials in a single neuron. 
Hodgkin-Huxley membrane model was proposed by Alan Lloyd Hodgkin and Andrew Fielding Huxley in 1952 (recipients of Nobel Prize in Physiology or Medicine 1963) to describe the ionic mechanisms of action potential in the squid's giant axon \cite{hodgkin1952components}.
Morris-Lecar model was developed by Catherine Morris and Harold Lecar in 1981 to reproduce the oscillatory behavior relevant to calcium and potassium conductance in the muscle fiber of the giant barnacle \cite{morris1981voltage}.
Fitzhugh-Nagumo model was suggested by Richard Fitzhugh in 1961 and created by Nagumo in the following year. Fitzhugh-Nagumo model is specialized in spike generation \cite{fitzhugh1961impulses,nagumo1962active}.
In all three models, equivalent circuits are utilized for modeling the excitable systems in neurons. Hodgkin-Huxley membrane model possesses the most complete form, whereas Morris-Lecar and Fitzhugh-Nagumo models are the simplified forms of Hodgkin-Huxley membrane model. 

The summation of the current at single neuron level is approximated as the measured EEG/MEG response in conductance-based neural mass models \cite{jones2007neural}.
The basic differential equations for one subneuronal population are 
\begin{align}
C\dot{V}(t) &= g(V_{rev}-V) + \Gamma, \\
\dot{g}(t) &= \kappa( \gamma \sigma (\mu_{off}-v_{threshold} \ , \ \Sigma_{off}) - g) + \Gamma.    
\end{align}
Capacitance $C$ times change of membrane potential $V$ is the equivalent current, which is dependent on conductance $g$, reverse potential $V_{rev}$, and unit noise $\Gamma$. The sigmoid function in convolution-based neural mass models becomes a cumulative distribution function for the normal distribution with afferent firing mean $\mu_{off}-V_{threshold}$ and firing variance $\Sigma_{off}$. $\kappa$ stands for the time constant, and $\gamma$ represents the strength of the connectivity \cite{moran2011consistent}.

\paragraph{\textbf{Neural Mass Model}} \hfill

NMM was coined by Marreiros et al. \cite{marreiros2010dynamic} for conductance-based neural mass model without the consideration of noise. That is, $\Gamma$ terms are removed in Equations (9) and (10) in NMM. NMM has been applied to examine distinction among the types of neurotransmitter receptors. In \cite{moran2011vivo}, the change of connectivity ($\gamma$) was examined for the effectiveness of pharmacologically induced changes in neurotransmitter receptors.

\paragraph{\textbf{Mean Field Model}} \hfill

Mean field model (MFM) differs from NMM by including the covariances of the observed time series data. Taking the covariances and noise into account, MFM can test the stability and capture the behaviors of quasi-periodic spikes. In \cite{marreiros2010dynamic}, Marreiros used MFM and reported that the variance of a simulated evoked potential elegantly coupled with the variance of the pyramidal cell's depolarization. 

\subsection{Field-based Neural Mass Model} \hfill
\\
Field-based neural mass models view the cortices as sheets and model the current as continuous flux on the cortical sheets. Each cortical column becomes a \textit{point} in the spatial field. Since the spatial domain is introduced into the time series neuronal signals, partial differential equations are employed in field-based neural mass models \cite{pinotsis2012dynamic}. Plugging \textit{field} concept can accommodate frequency dependent spatially propagated activities on cortical sheets.

The basic differential equation for one subneuronal population in neural field accommodation is
\begin{align}
\ddot{v}+2 \kappa_{e/i} \dot{v} + \kappa_{e/i} v &= \kappa_{e/i} H_{e/i} \int \int D(x,t) \sigma(v) dx \ dt .    
\end{align}
Potential $v$ is spatially and temporally dependent. $\kappa$ is the inverse of time constant, and $H$ is the magnitude of maximum neuronal post-synaptic potentials. Connectivity matrix $D$ contains the message passing from both time and spatial domains. $\sigma$ is the sigmoid function. The differential equation has similar structure as the second order equation (Equation (5) ) proposed in Jansen and Rit's paper \cite{jansen1995electroencephalogram} except that the input transforms into the integral of spatially propagated signals.

\paragraph{\textbf{Neural Field Model}} \hfill
\\
Neural field model (NFM) shares similar structure as convolution-based neural mass model except the input is spatially dependent. The invention of NFM leads to the construction of canonical micro-circuits. Pitnosis et al. applied NFM and successfully established the correlation between peak gamma frequency and the size of visual cortex found in Muthukumaraswamy et al.'s study \cite{pinotsis2013dynamic,muthukumaraswamy2009resting} by adding one more subneuronal population (four in total).  

\paragraph{\textbf{Canonical Micro-Circuits}} \hfill
\\
Canonical micro-circuits (CMC) differs from NFM with an additional subneuronal population at supra-granular layer. CMC is based on the intra-cellular recordings in cats' visual cortex conducted by Douglas and Martin in 1991 \cite{douglas1991functional}. Currently, CMC are utilized to test the circuits/connectivites at GABAergic interneurons (inhibitory interneurons layer) \cite{packer2011dense}.


\begin{table}[!htbp]
\centering
\begin{tabularx}{\textwidth}{ |X|X|X|X| }
  \hline
  Models & Applications & Advantages & Disadvantages \\
  \hline 
  Event Related Potential (convolution-based)  & Mismatch negativity & Simple but sufficient to generate alpha activity & Cortical region assumed as a point mass \\
  \hline
  Local Field Potential (convolution-based) & Synaptic function & Regulation of inhibitory interneuron in addition to ERP model & Cortical region assumed as a point mass \\
  \hline
  Neural Mass Model (conductance-based) & Distinct the types of neurotransmitter receptors & Reasonable biological structure & Relatively many parameters to estimate \\
  \hline
  Mean Field Model (conductance-based) & Behaviors of quasi-periodic spikes & High order statistics included & Relatively many parameters to estimate \\
  \hline
  Neural Field Potential (field-based) & Spectral Analysis  & Spatial distance considered & The whole brain assumed as a field \\
  \hline
  Canonical Micro-Circuits (field-based) & Connectivities tested at GABAergic interneurons & Additional subneuronal population & The whole brain assumed as a field \\
  \hline
\end{tabularx}
\caption{Table of Neuronal Population Models' Advantages / Disadvantages}
\label{tab:npm}
\end{table}

\subsubsection{\textbf{Challenges}} \hfill 
\\
We summarize the applications of each model in Table \ref{tab:npm}. ERP model has been applied to study mismatch negativity, and ERP model is the simplest among the six models. LFP has been employed for investigating synaptic functions, and compared to ERP, LFP has additional regulation of inhibitory interneuron. Both ERP and LFP belong to convolution-based neural mass models, which can be too simplistic because each neuronal population is assumed as a point of mass. NMM has been utilized to distinguish different types of neurotransmitters; MFM has been adopted to study the behaviors of quasi-periodic spikes. Both NMM and MFM are established on the neuronal cellular structures, and thus, provide more biologically reasonable information. However, the relatively many parameters to estimate (compared to ERP and LFP) can sometimes lead to more challenging parameter estimation tasks. NFM has been conducted on spectral analysis, and CMC has been applied on testing the connectivities at GABAergic interneurons. Both NFM and CMC belong to field-based neural mass models, Spatial distance is considered in both NFM and CMC, but the calculation of signal propagation may be biased due to the field assumption. The models with more parameters, for example, conductance-based to convolution-based neural mass models, provide more neurological information and biological interpretations. The relatively simple models are more robust and the fine-tuned parameters are more convincing. A method to select the optimal model for comparing the results or testing the hypothesis has not yet been established. This highlights the need for an approach to select the appropriate model for the suitable case.

\section{Effective Networks Estimation}
\label{sec:ENN}


In this report, three state-of-the-art effective networks estimation methods are surveyed: dynamic causal modeling, structural causal modeling, and vector autoregression.

\subsection{Dynamic Causal Model} \hfill
\\
Dynamic causal model (DCM) employs parameter estimation method and performs model comparison to fine-tune the connectivity parameters $\Theta$ in Equations (1) and (2) for causal inference \cite{friston2003dynamic}. Currently, mean field variational Bayesian approach is adopted in DCM to estimate the effective network. Model comparison is completed by computing the logarithm of evidence ratio
$$ln(\frac{Pr(y|m_1)}{Pr(y|m_0)})=ln(Pr(y|m_1))-ln(Pr(y|m_0)),$$
in which $Pr(y|m,u) = \int Pr(y,x,\Theta|m,u) dx d\Theta$ is the model evidence. Given its flexibility in model selection and adaptability in data types, DCM has gradually gained increasing usage for effective connectivity estimation of fMRI, EEG, and MEG data \cite{schuyler2010dynamic,friston2011network,brown2012dynamic,adams2016dynamic}. The formulation in DCM for parameter estimation (refer to state-space transition equations in Equations (1) and (2) ) is
\begin{equation}\label{eq:bilinearApp}
    \dot{x}(t) = ( A + \Sigma_{j=0}^{J} B^{j} u_j ) x(t) + C u(t) +w(t),
\end{equation}
where $A = \frac{\partial f}{\partial x}$, $B=\frac{\partial^2 f}{\partial x \partial u}$, and $C=\frac{\partial f}{\partial u}$. Letter $j$ denotes the $j^{th}$ input to the system. Matrix $A$ is the Jacobian matrix; matrix $B$ encodes the change of state variable $x$ generated by input $u$; matrix $C$ contains the direct influence from input $u$. The strength of DCM is to integrate differential equations into causal relationship inference. However, variational Bayesian technique utilized in DCM frequently gets trapped at local optimum, and model selection approach might be subjective due to the choices of the available models.

\subsection{Structural Causal Model} \hfill
\\
Strucutral causal model (SCM) rests on structure equation modeling, which assumes the equilibrium is reached at each point of time series data. SCM is at first a generic approach designed for solving social and economics problems. In accord with SCM, Equations (1) and (2) are reduced to
\begin{equation}
    0 = Ay(t) + w(t), 
\end{equation}
with the constraints of $y(t) = x(t)$, $\dot{x}(t) = 0$, and $u(t)=0$. SCM has been put into research in seeking the association between the task and the active regions of fMRI studies. For example, bilateral primary motor cortices have been discovered to correlate with finger tapping frequency temporally in Zhuang et al.'s study \cite{zhuang2005connectivity}. SCM has relatively simple formulation than DCM does, and the inference made by SCM is more robust compared to DCM. The drawback of SCM is its disability to capture the properties of fast-changing or nonstationary neuronal signals. 

\subsection{Vector Autoregression} \hfill
\\
Vector autoregression (VAR) method relies on temporal precedence to make inference on causal relation. In neuroscience, the causality inferred from VAR is known as directed functional causality (not effective causality). Since VAR depends directly to the time series observations, the formulation for VAR can be written as
\begin{equation}
    y(t) = \tilde{A} x(t-\delta) + z(t),
\end{equation}
where $\tilde{A}=exp(\delta A)$, $A$ is the Jacobian matrix, and $z(t) = \int_{0}^{\delta} exp(\tau A)w(t-\tau) d\tau $. VAR has been applied in many research works aiming at making causality inference from neuroimaging data \cite{goebel2003investigating,dhamala2018granger,sheikhattar2018extracting}. The famous (Geweke) Granger causality belongs to VAR approach, and Granger causality alone cannot describe the underlying interaction among neuronal populations that complies with the biophysical models. Therefore, techniques developed for dynamic Granger causality are often employed in fMRI studies. Nevertheless, when the signal to noise ratio is low, VAR suffers from generating reliable causality inference.

\begin{table}[!htbp]
\begin{center}
\begin{tabularx}{\textwidth}{ |X|X|X|X| }
  \hline
  Models & Advantages & Disadvantages \\
  \hline 
  Dynamic Causal Model (DCM) & \begin{flushleft}
  \begin{itemize}
  \item Biologically reasonable
  \item Causal relationship included
  \end{itemize}
  \end{flushleft} & \begin{flushleft}
  \begin{itemize}
  \item Local optima found 
  \item Relatively complicated derivation
  \end{itemize}
  \end{flushleft} \\
  \hline
  Structural Causal Model (SCM) & \begin{flushleft}
  \begin{itemize}
  \item Robust compared to DCM
  \item Able to test correlations between tasks and active brain regions
  \end{itemize}
  \end{flushleft}  & \begin{flushleft}
  \begin{itemize}
  \item Unable to apply to fast-changing signals
  \item Biological state space equations not considered
  \end{itemize}
  \end{flushleft} \\
  \hline
  Vector Autoregression (VAR) & \begin{flushleft}
  \begin{itemize}
  \item Geweke Granger causality included
  \item Applied in many control theory related studies
  \end{itemize}
  \end{flushleft}  & \begin{flushleft}
  \begin{itemize}
  \item Unable to generate reliable causality inference when signal to noise ratio is low
  \item Biological state space equations not considered
  \end{itemize}
  \end{flushleft}  \\
  \hline
\end{tabularx}
\end{center}
\caption{Table of Effective Network Methods' Advantages / Disadvantages}
\label{tab:enem}
\end{table}

\subsection{\textbf{Challenges}} 

We encapsulate the merits and drawbacks of each network estimation approach in Table \ref{tab:enem}. DCM merges the constraints of biological models into effective network estimation. Such strategy is favorable to neuroscientists but the estimated networks may not be optimal. SCM assumes the equilibrium is reached in the measured signals. SCM is more robust than DCM but unable to apply to fast-changing biological signals. VAR employs Geweke Granger causality and has been applied to several control related territories. VAR, however, does not consider biological state space equations, so the estimated networks are sometimes difficult to interpret. The properties of the three surveyed methods indicate the need of an effective network estimation approach that is not only robust but also takes state transition in biological models into account.

\section{Dynamic Parameter Estimation Methods}
\label{sec:estimation}

In this section, we survey dynamic parameter estimation techniques for nonlinear systems from three perspectives: physics, Bayesian statistics, and optimization. First, variational Bayesian (VB) inference method, which is categorized into physicists' viewpoint in this report, is introduced. Next, simulated annealing method such as Markov Chain Monte Carlo is elaborated as strategies taken by Bayesians. Last, constrained optimization technique, which is a nonlinear programming method, is unrolled. Numerous methods can be discussed in inverse problems/reverse engineering, but in this report, we only cover the selected nonlinear system parameter estimation methods.\\

\noindent \textbf{Notations}\\
In the following paragraphs of this section, the targeted parameters, strengths/delays/connectivities, are denoted as $\theta$; feasible data's states, potentials/currents of neuronal populations, are represented as $x$; infeasible data's states, hidden variables, are symbolized as $z$ and neuroimaging data as $y$. 

\subsection{Perspective From Physics - Variational Bayesian Approach}
In this subsection, we first introduce Bayes' Theorem and expectation maximization (EM) method, and then discuss Variational Bayesian (VB) method.

Bayes' Theorem describes the probability of an event under the constrained knowledge or priors \cite{sivia2006data}, and is expressed as:
\begin{equation}\label{eq:bayes}
Pr(\theta|y) = \frac{Pr(y|\theta)Pr(\theta)}{Pr(y)}.
\end{equation}
In Equation (\ref{eq:bayes}), $Pr(\theta)$ is the prior distribution, which is subjective to the observers' choices; $Pr(y|\theta)$ is the likelihood function, which is a conditional probability distribution; $Pr(y)$ is the probability distribution of observing the data $y$; $Pr(\theta|y)$ is the posterior distribution, which is the probability distribution of the parameters conditioned on the data. 

In neuroscience parameter estimation problem, the goal is to seek the parameters of neuronal populations that maximize the probability of the posterior distribution given the neuroimaging data $y$:
\begin{equation}
\argmax_{\theta} = Pr(\theta | y).
\end{equation}
In application, direct calculation of the posterior distribution is often infeasible due to the introduction of hidden (latent) variables ($Z$). Therefore, solving the equation directly is completely difficult, but an intelligent approach, EM algorithm, was proposed to handle such issue. 

In EM algorithm [\ref{alg:EM}], two steps are executed iteratively until the convergence criterion is met.
One step is called the E-step, and in E-step, the expected value of the hidden variable(s) $Z$ is computed based on the given parameters. The other step is M-step, and in M-step, the parameters that maximize the posterior distribution are calculated. Throughout the iterations, the initialized parameters will converge toward the parameters that maximize the marginal likelihood of the observed data. However, the calculation of the probability distribution of hidden variable $Pr(Z|y,\theta)$ is often intractable, so VB method is introduced to solve the intractable problem \cite{tzikas2008variational}.

\begin{algorithm}[H]
\caption{Expectation-Maximization Algorithm}
\label{alg:EM}
\SetAlgoNoLine
Input: parameter $\boldsymbol{\vartheta^{0}}$, observation $\boldsymbol{y}$ \\
Output: parameter $\boldsymbol{\vartheta}$\\
Initialize $\boldsymbol{\vartheta}=\boldsymbol{\vartheta}^0$ \\
$N\_epoch=$ iteration upper bound \\
\While{$i<N\_epoch$}{
\Repeat{convergence}{
{\bf E-step}\\
\hspace{1cm} $Q(\boldsymbol{\vartheta | \vartheta^{i} }) =  \boldsymbol{E}_{\boldsymbol{Z|y,\vartheta}} \ln (Pr(\boldsymbol{Z|y,\vartheta}))$  \\
{\bf M-step}\\
\hspace{1cm} $\boldsymbol{\vartheta^{i+1}} = \argmax_{\boldsymbol{\vartheta} } Q( \boldsymbol{\vartheta | \vartheta^{i} }) $
}
}
\end{algorithm}
VB suggests a known function $q(z)$ to approximate the intractable function $ \ln (Pr(\boldsymbol{Z|y,\vartheta}))$. Among VB techniques, mean field approximation is frequently used in statistical physics. Under the accommodation of mean field approach, the objective function is reformulated as the minimization of Kullback-Leibler divergence of the proposed distribution $q(\theta)$ and conditional probability distribution $Pr(y|\theta)$:
\begin{align}
min \quad KL(q(z)\|Pr(y|\theta) ),
\end{align}
subject to
\begin{flalign*}
q(z) &= h(\theta) \\
p(\theta) &= p(\mu) p(\sigma) \\
p(\mu) &= N(\Lambda^\mu,\lambda^\mu) \\
p(\sigma) &= N(\Lambda^\sigma,\lambda^\sigma)
\end{flalign*}
in which $q(z)$ is the probability distribution of the hidden variables; probability distribution $p(\theta)$ is assumed to be Gaussian distribution; function $h$ transforms the parameters to the potentials/currents of the neuronal populations. In fact, minimizing Kullback-Leibler divergence $KL(q(z)||Pr(y|\theta))$ is equivalent to maximizing the free energy $F$, and the new objective function can be formulated as:
$$max \quad E_{z \sim q(z|y,\theta)} log (Pr(y|z,\theta) ) - KL(q(z|y,\theta)\|Pr(z| \theta)).$$

The VB version EM algorithm [\ref{alg:VBEM}] then becomes \cite{tzikas2008variational}:

\begin{algorithm}[H]
\caption{Variational Expectation-Maximization Algorithm}
\label{alg:VBEM}
\SetAlgoNoLine
Input: parameter $\boldsymbol{\vartheta^{0}}$, distribution $\boldsymbol{q}$ \\
Output: parameter $\boldsymbol{\vartheta}$\\
Initialize $\boldsymbol{\vartheta}=\boldsymbol{\vartheta}^0$ \\
$N\_epoch=$ iteration upper bound \\
\While{$i<N\_epoch$}{
\Repeat{convergence}{
{\bf Variational E-step}\\
\hspace{1cm} Evaluate $q^{i+1} (z)$ to maximize $F$\\
{\bf Variational M-step}\\
\hspace{1cm} $\boldsymbol{\theta^{i+1}} = \argmax_{\boldsymbol{\theta} } F $
}
}
\end{algorithm}

\medskip

\noindent \textbf{From the perspective of physics:} Distribution $q(z)$ is updated to minimize the variational free energy $F(q,\theta)$. The free energy represents the divergence between the real and approximate conditional density minus the log-likelihood. In M-step of VB EM algorithm, parameters $\theta$ are updated to minimize the discrepancy between the true and approximate conditional density, which is equivalent to maximize the log likelihood. Once the posterior density $q(z)$ is determined, the inference on the parameters of a particular model can be specified.

\subsection{Perspectives From Bayesian Statistics - Simulation Techniques }

Rejection sampling, importance sampling, particle filtering, and Markov Chain Monte Carlo (MCMC) are covered in this section.
All of the aforementioned methods employ the idea of Monte Carlo inference and attempt to overcome the intractability of the posterior distribution - in a simulation tactic. Through simulations, the posterior can be approximated via the computation of the largely generated samples.

\subsubsection{Rejection Sampling and Importance Sampling} \hfill
\\
In rejection sampling [\ref{alg:rejectsample}], users propose a distribution $q(x)$ from which samples are drawn, and samples are accepted if $\frac{p(x)}{q(x)}>K$, in which $K$ is a value chosen by the user. While in importance sampling, one would like to sample $x$ in high probability regions, plus in the regions where $|f(x)|$ is large.

\begin{algorithm}[H]
\caption{Single Sample acquisition of Rejection Sampling Algorithm}
\label{alg:rejectsample}
\SetAlgoNoLine
Input: distributions $p$, $q$ \\
Output: samples $\boldsymbol{x}$ \\
$M=$ sampling times for one sample \\
$K$ = acceptance value \\
\While{$k<M$}{
\Repeat{}{
Draw a sample $\boldsymbol{x}^{*}$ from the distribution $q(\boldsymbol{x})$ \\
Calculate $ \alpha = \frac{p(x)}{q(x)}$ \\
Accept $\boldsymbol{x}^{*}$ if $alpha > K $; otherwise, reject $\boldsymbol{x}^{*}$ 
}

}

\end{algorithm}

\subsubsection{Particle Filtering} \hfill
\\

Particle filtering is a sequential Monte Carlo approach, and users applying such approach take interest in the state-space transition. Particle filtering is often used in nonlinear/non-analytic dynamic equations. The state and observed data's update equations for our application are written as:
\begin{align}
X_{t+1} &= f(X_t,u_{t+1},\theta_{t+1}),\\
Y_{t+1} &= g(X_t,\theta_{t+1}),
\end{align}
How particle filtering works can be analogous to solving the problem of figuring out the location of a driver, who only has a map to refer to. Initially, any location on the map is possible for the driver to be. Next, throughout driving a certain amount of distance, the driver gradually filters out the impossible locations, based on the driving path. Eventually, the driver can locate his/her position on the map by adopting the filtering process.

\begin{algorithm}[H]
\caption{Simple Version of Particle Filtering Algorithm}
\label{alg:PF}
\SetAlgoNoLine
Input: probability distributions $Pr(\theta)$, $Pr(\theta,X|y)$ \\
Output: samples $\boldsymbol{x}$ \\
$T=$ sequential time length \\
\textbf{Initialize} state $\boldsymbol{X_0}, \quad t=1$ \\
\While{$t<T$}{
\Repeat{t=T}{
{\bf Step 1. Prediction: } Draw N samples \{ $X^{t|t-1,k}, \theta^{t|t-1,k}$ \} from the conditional density $Pr(\theta_t) \, Pr(\theta^{t-1},X_0|y^{t-1})$\\
{\bf Step 2. Filtering: } Assign the weighting to each draw from \textbf{Step 1} as $\frac{Pr(y_t|\theta^{t|t-1,k},x_0^{t|t-1,k},y^{t-1})}{\sum_{k=1}^{N} Pr(y_t|\theta^{t|t-1,k},x_0^{t|t-1,k},y^{t-1})}$ for each draw  $x^{t|t-1,k}, \theta^{t|t-1,k}$ \\
{\bf Step 3. Sampling: } Draw N samples again based on the filtering step's weighting assignment, and replace the N samples in Step 1 with the new samples
}
}

\end{algorithm}

\subsubsection{Markov Chain Monte Carlo} \hfill
\\

Last, we introduce MCMC method. MCMC shares the common goal with EM method by easing the computation of the posterior distribution, and the core concept of MCMC is to estimate $\int P(Y|\boldsymbol{\vartheta})P(\boldsymbol{\vartheta}) \, d \boldsymbol{\vartheta}$ precisely by simulations through sampling. Estimated parameters are often selected to be normal distributions, and under such assumption, the distributions of the model become
\begin{equation}
\boldsymbol{\vartheta} \sim P(\boldsymbol{\vartheta}), \quad y_{ji}|\boldsymbol{\vartheta} \sim N(\boldsymbol{x(t_j,\vartheta)},\sigma_i^2),
\end{equation}
where $y_{ji}|\boldsymbol{\vartheta}$ indicates a conditional distribution. Given the prior and the conditional distributions, the calculation of a posterior probability distribution of $\boldsymbol{\vartheta}$ is
\begin{equation}
P(\boldsymbol{\vartheta}|Y) = K^{-1} P(Y|\boldsymbol{\vartheta})P(\boldsymbol{\vartheta}) = K^{-1}e^{l(\boldsymbol{\vartheta})}P(\boldsymbol{\vartheta}), \, l(\boldsymbol{\vartheta}) = log P(Y|\boldsymbol{\vartheta}),
\end{equation}
where $K$ is a normalizing constant, $K = \int P(Y|\boldsymbol{\vartheta})P(\boldsymbol{\vartheta}) \, d \boldsymbol{\vartheta} $. Once the posterior distribution is obtained, one can estimate the parameters from the expected values
\begin{equation}
\tilde{\boldsymbol{\vartheta}} = \int \boldsymbol{\vartheta} P(\boldsymbol{\vartheta}|Y) \, d \boldsymbol{\vartheta}.
\end{equation}
Normalizing constant $K$ can be estimated precisely to a certain degree once sufficiently large samples from $\vartheta_1, \cdots , \vartheta_M$ are gathered. A simple MCMC takes two steps: the first is to draw a sample from $\boldsymbol{\vartheta}$, which is a proposal of a move from current state to the next $\vartheta^k \rightarrow \vartheta^{k+1}$; then, one accepts that step and make a move, or rejects it and stays at $\vartheta^k$. \\

\begin{algorithm}[H]
\caption{Simple Version of Markov Chain Monte Carlo Algorithm}
\label{alg:MCMC}
\SetAlgoNoLine
Input: distributions $P$, $Q$ \\
Output: parameter $\boldsymbol{\vartheta}$ \\
$N\_epoch=$ sample times upper bound \\
\While{$k<N\_epoch$}{
\Repeat{convergence}{
Draw a sample $\boldsymbol{\vartheta}^{*}$ from the proposed distribution $Q(\boldsymbol{\vartheta}^{*}|\boldsymbol{\vartheta}^{k})$ // based on neuroimaging data observed \\
Calculate $ \alpha = \frac{P(Y|\boldsymbol{\vartheta}^{*}) P(\boldsymbol{\vartheta}^{*}) Q(\boldsymbol{\vartheta}^{k}|\boldsymbol{\vartheta}^{*}) }{P(Y|\boldsymbol{\vartheta}^{k}) P(\boldsymbol{\vartheta}^{k}) Q(\boldsymbol{\vartheta}^{*}|\boldsymbol{\vartheta}^{k})}$ \\
Accept $\boldsymbol{\vartheta}^{*}$ and set $\boldsymbol{\vartheta}^{k+1}=\boldsymbol{\vartheta}^{*}$ with probability $\alpha$; otherwise, reject $\boldsymbol{\vartheta}^{*}$ and set $\boldsymbol{\vartheta}^{k+1}=\boldsymbol{\vartheta}^{k}$
}
}

\end{algorithm}
\noindent In the acceptance probability $\alpha$, the ratio $\frac{P(Y|\boldsymbol{\vartheta}^{*}) }{P(Y|\boldsymbol{\vartheta}^{k})}$ measures the density distribution of $\boldsymbol{\vartheta}^{k+1}$ to $\boldsymbol{\vartheta}^k$, while $\frac{P(\boldsymbol{\vartheta}^{*}) }{ P(\boldsymbol{\vartheta}^{k})}$ balances the probability of moving from $\boldsymbol{\vartheta}^k$ to $\boldsymbol{\vartheta}^{k+1}$. When the targeted value is approximated, the samples are trapped in their stationary distributions. Transition distribution  $Q(\boldsymbol{\vartheta}^{*}|\boldsymbol{\vartheta}^{k})$ is usually implemented with the normal distribution:
\begin{equation}
\boldsymbol{\vartheta}^{*} = \boldsymbol{\vartheta}^k + \boldsymbol{\epsilon}, \boldsymbol{\epsilon} \sim N(0, \tau^2 \boldsymbol{I}).  
\end{equation}
Empirically, the selection of the variance is important, and good results often take on $\alpha$ being around $0.25 \sim 0.3$.

\medskip

\noindent \textbf{Perspective from Bayesian statistics} agrees that the approximation of a true distribution can be calculated with sufficiently large samples probabilistically. A good proposed distribution can accelerate the computation of the target; however, the convergence of the calculated result is often not guaranteed in highly complicated kinetic models. Furthermore, the scalability and dimensions can slow down the simulation time.
\subsection{Constrained Optimization: Gauss-Newton Algorithm with Collocation Method}

In this section, we discuss a constrained optimization approach by first covering the essential materials of the Gauss-Newton method, which is the oldest and still the most popular Nonlinear Least Squares (NLS) approach, and the collocation method. Then, we show how to use the hybrid of the two materials to formulate the constrained optimization problem.


\begin{center}
{\bf Gauss-Newton Algorithm}
\end{center}

In this subsection, we go over the application of Gauss-Newton algorithm [\ref{alg:GN}] onto a standard ordinary differential equation (ODE) with initial value problem, including the introduction of sensitivity equations, extension to a multivariate system, and practical problems in using Gauss-Newton algorithms. The modeled problem is
\begin{equation}
\frac{d}{dt} \mathbf{x}(t) = \mathbf{f}  ( \mathbf{x}(t),u(t), \boldsymbol{\theta} ) \quad and \quad \boldsymbol{x}(t_0) = \boldsymbol{x}_{0},
\end{equation}
where vector $\boldsymbol{x}(t)$ contains the potentials/currents of neuronal population.
We use $g^{-1}(y)$ to denote the feasible $\boldsymbol{x}(t)$. As for the infeasible potentials/currents, we can do the estimation via forward calculation of the neural mass model with the estimated parameters. Often, the initial state is also unknown, and can also be categorized into the estimated term. Therefore, the \textit{var theta} symbol $\vartheta$, instead of $\boldsymbol{\theta}$, is applied to indicate the parameter subset $ \{ \boldsymbol{\theta},\boldsymbol{x}_0 \} $. Suppose a univariate observation $g^{-1}(y)$ is provided, without loss of generality, the observational error at time $t_j$ with the corresponding state variable $x$ is
\begin{equation}
\epsilon_j = g^{-1}(y_j) - x(t_j,\boldsymbol{\vartheta}).
\end{equation}
Abiding by the sum of the squared errors (SSE) criterion to nonlinear regression, the best-fitted parameters should generate the least errors
\begin{equation}
\argmin_{\boldsymbol{\vartheta}} = \sum_{j=1}^{n}{(g^{-1}(y_j)-x(t_j,\boldsymbol{\vartheta}))}^2,
\end{equation}
where SSE is defined as
\begin{equation}
SSE(\boldsymbol{\vartheta}) = \sum_{j=1}^{n}{(g^{-1}(y_j)-x(t_j,\boldsymbol{\vartheta}))}^2.
\end{equation}

To minimize $SSE$, Gauss-Newton algorithm first makes a guess on initial parameters, and then iterates through the search of a better-fitted parameters with gradient descent technique until the convergence criterion is met. The initial parameters are denoted as $\boldsymbol{\vartheta}^0$. The next step is to do Taylor expansion at $\boldsymbol{\vartheta}^0$ to second order (Equation 28)
\begin{equation}
SSE( \boldsymbol{\vartheta}) \approx SSE(\boldsymbol{\vartheta}^0) + \partial_{\boldsymbol{\vartheta}}SSE(\boldsymbol{\vartheta}^0)(\boldsymbol{\vartheta}-\boldsymbol{\vartheta}^0)+ \frac{1}{2}(\boldsymbol{\vartheta}-\boldsymbol{\vartheta}^0)^T \partial_{\boldsymbol{\vartheta}}^{2}SSE(\boldsymbol{\vartheta}^0)(\boldsymbol{\vartheta}-\boldsymbol{\vartheta}^0)
\end{equation}
\begin{align}
\partial_{\boldsymbol{\vartheta}}SSE(\boldsymbol{\vartheta}^0) &= -2 \sum_{j=1}^{n} \partial_{\boldsymbol{\vartheta}}x(t_j,\boldsymbol{\vartheta}) (g^{-1}(y_j) - x(t_j,\boldsymbol{\vartheta})) = -2\partial_{\boldsymbol{\vartheta}}\boldsymbol{x(\vartheta)}^T \boldsymbol{(g^{-1}(y)-x(\vartheta))} \\
\boldsymbol{J(\vartheta)} &= \big[ \boldsymbol{J(\vartheta)} \big]_{jl} = \frac{dx(t_j,\boldsymbol{\vartheta})}{d \vartheta_l} \\
\partial_{\boldsymbol{\vartheta}}^{2}SSE(\boldsymbol{\vartheta}^0) &=-2 \boldsymbol{J(\vartheta)}^T \boldsymbol{J(\vartheta)}-2 \partial_{\boldsymbol{\vartheta}}^{2} x(\boldsymbol{\vartheta})^T (\boldsymbol{g^{-1}(y)-x(\vartheta)})
\end{align}
where $\partial_{\boldsymbol{\vartheta}}SSE(\boldsymbol{\vartheta}^0)$ is the \textit{gradient vector} at the initial guess, $\boldsymbol{J(\vartheta)}$ is the \textit{Jacobian matrix}, and $\partial_{\boldsymbol{\vartheta}}^{2}SSE(\boldsymbol{\vartheta}^0)$ is the corresponding \textit{Hessian matrix}. Omitting $\partial_{\boldsymbol{\vartheta}}^{2} x(\boldsymbol{\vartheta})^T (\boldsymbol{g^{-1}(y)-x(\vartheta)})$ will lead to minor perturbation, largely reduced computation cost, and an invertible Hessian matrix expression. Hence, only the first term of the complete Hessian matrix will be used. To minimize SSE at $\boldsymbol{\vartheta}^0$, by doing the differentiation derivation, the equation becomes
\begin{equation}
\boldsymbol{\vartheta}^1 = \boldsymbol{\vartheta}^0- \big[ \partial_{\boldsymbol{\vartheta}}^2 SSE(\boldsymbol{\vartheta}^0) \big]^{-1} \partial_{\boldsymbol{\vartheta}}SSE(\boldsymbol{\vartheta}^0).
\end{equation}

\begin{algorithm}[H]
\caption{Gauss-Newton Algorithm}
\label{alg:GN}
\SetAlgoNoLine
Input: Observation $\boldsymbol{y}$, observation function $\boldsymbol{g}$, differential equations $\boldsymbol{f}$ \\
Output: Parameters $\boldsymbol{\vartheta}$\\
Initialize $\boldsymbol{\vartheta}=\boldsymbol{\vartheta}^0$ \\
$N\_epoch=$ iteration upper bound \\
\While{$i<N\_epoch$}{
\Repeat{convergence}{
$\boldsymbol{H(\vartheta^k)} = \boldsymbol{J(\vartheta^k)^TJ(\vartheta^k)}$ \\
$\boldsymbol{g(\vartheta^k)} = \boldsymbol{J(\vartheta^k)^T (g^{-1}(y)-x(\vartheta^k))}$ \\
$\boldsymbol{(\vartheta^{k+1})} = \boldsymbol{\vartheta^k + H(\vartheta^k)^{-1} g(\vartheta^k)}$
}
}

\end{algorithm}

\vspace{20mm}

\noindent {\bf Sensitivity Equations}\\

Solving the sensitivity equations makes Gauss-Newton algorithm plausible because computing the Jacobian matrix  $\boldsymbol{J(\vartheta)}$ is usually difficult in Gauss-Newton algorithm's implementation. Once the initial conditions are set, the target is reachable by implicit differentiation:
\begin{gather}
 \frac{d}{dt} \begin{bmatrix} x  \\ \partial_{\theta} x \\ \partial_{x_0}x  \end{bmatrix}
 =
  \begin{bmatrix}
   f ( x(t;\boldsymbol{\theta},x_0),u,\boldsymbol{\theta}) \\
   \partial_{\theta} f (  x(t;\boldsymbol{\theta},x_0),u ,\boldsymbol{\theta}) + \partial_{x} f (  x(t;\boldsymbol{\theta},x_0),u ,\boldsymbol{\theta}) \, \partial_{\theta} x(t;\boldsymbol{\theta},x_0) \\
   \partial_{x} f (  x(t;\boldsymbol{\theta},x_0),u ,\boldsymbol{\theta}) \, \partial_{x_0} x(t;\boldsymbol{\theta},x_0) 
   \end{bmatrix}
\end{gather}
with the corresponding initial conditions
\begin{gather}
 \frac{d}{dt} \begin{bmatrix} x(t_0)  \\ \partial_{\theta} x(t_0) \\ \partial_{x_0}x(t_0)  \end{bmatrix}
 =
  \begin{bmatrix}
   x_0 \\
   \boldsymbol{0} \\
   \boldsymbol{I}
   \end{bmatrix}
\end{gather}

\medskip

\noindent {\bf Measurements on Multiple Variables}\\

Different from a single variable, multivariate Gauss-Newton method is an extension of univariate Gauss-Newton method. Most of the procedures are similar, but the users are allowed to provide weights for each variable. Different weights in a multivariate system is reasonable since each variable accounts for different scales and measurement precision.


\medskip

\noindent{\bf Practical Problems in Gauss-Newton Methods}\\

\begin{itemize}
\item \textbf{Local Minima:} The hyper-surfaces of $SSE$ may contain a bunch of local minima. Our goal is to seek the global minimum, and the Gauss-Newton method can get trapped at local minima.
\item \textbf{Initial Parameter Values:} Since the finding of global minimum is not guaranteed by using the Gauss-Newton method, the choice of initial parameter values becomes important. If a good initialization is made, the estimated parameters are close to the real values.
\item \textbf{Identifiability:} In dynamic parameter estimation, parameter identifiability is an important issue. In highly dimensional system, some parameters, or some combinations of parameters are nearly unidentifiable. One phenomenon observed by \cite{brown2004statistical} and \cite{gutenkunst2007universally} shows the eigenvalues of the Hessian matrix tends to decay exponentially when the data are very informative.
\end{itemize}

\begin{center}
{\bf Collocation Methods}
\end{center}

Collocation method is an approach to find numerical solutions in differential equations or integral equations mathematically; it constructs the space with a finite number of bases, and the equations are solved at the collocated points. Such method explicitly represent $\boldsymbol{x}(t)$ as a linear combination of a set of (predefined) basis functions:
\begin{equation}
\boldsymbol{x}(t) \approx \sum_{k=1}^{K} \boldsymbol{c}_k \phi_k (t) = \boldsymbol{C \phi}(t),
\end{equation}
where $\phi_k(t)$ are a pre-chosen set of functions and weights $\boldsymbol{c}_k$ correspond to each basis function. $\boldsymbol{\phi}(t)$ is the matrix composed of $\phi_k(t)$ at the $k^{th}$ row, and $\boldsymbol{C}$ is a $1 \times K$ matrix with $c_k$ at the $k^{th}$ column. Following such expression, the derivatives of $\boldsymbol{x}(t)$ is:
\begin{equation}
\frac{d}{dt} \boldsymbol{x}(t) \approx \sum_{k=1}^{K} \boldsymbol{c}_k \frac{d}{dt} \phi_k (t). 
\end{equation}
\bigskip

\begin{center}
 {\bf Constrained Optimization using Collocation Methods}
\end{center}

Several parameter estimation methods are discussed in this part because many flexible changes can be made by tweaking the objective function. First, we introduce the formulation of the constrained optimization problem. Next, we discuss different parameter estimation methods including trajectory matching, gradient matching, smoothing methods, and profiled estimation method.\\

Combining collocation method and the SSE criterion together, the optimization problem is reformulated as:
\begin{equation} \label{eq:trajectorySSE}
(\boldsymbol{\vartheta,c_1, \cdots,c_d}) = \argmin \sum_{i=1}^{d}  \sum_{j=1}^{n} (g^{-1}( y_{ji} ) -\boldsymbol{\phi}(t)^T \boldsymbol{c}_i ) ^2 
\end{equation}
subject to
\begin{align}
\boldsymbol{\phi}(t_0) \boldsymbol{C} &= \boldsymbol{x}_0 \\
\frac{d}{dt} \boldsymbol{\phi}(t_l) \boldsymbol{C} &= \boldsymbol{f}(\boldsymbol{\phi}(t_l) \boldsymbol{C},u(t_l),\boldsymbol{\vartheta}) 
\end{align}

\noindent In dual space, we consider the Lagrangian: 
\begin{equation}
\boldsymbol{\Lambda(C,\vartheta,\lambda)} = \sum_{i=1}^{d}  \sum_{j=1}^{n} ( g^{-1} ( y_{ji} )-\boldsymbol{\phi}(t)^T \boldsymbol{c}_i ) ^2 + \boldsymbol{\lambda_0^T}[\boldsymbol{\phi}(t_0) \boldsymbol{C}-\boldsymbol{x}_0] + \sum_{l=1}^{K-1} \boldsymbol{\lambda}_l^T [\frac{d}{dt} \boldsymbol{\phi}(t_l) \boldsymbol{C} - \boldsymbol{f}(\boldsymbol{\phi}(t_l) \boldsymbol{C},u(t_l),\boldsymbol{\vartheta})].
\end{equation}
Suppose the optimal values exist for $\boldsymbol{\vartheta, \ C}$, and given the optimization problem has differentiable objective function and constraints, \textit{Karush-Kuhn-Tucker} (KKT) conditions must be satisfied at the optimal values $\boldsymbol{C^*,\vartheta^*}$ \cite{boyd2004convex}. Thus, the KKT conditions are expressed as 
\begin{align}
\boldsymbol{\phi}(t_0) \boldsymbol{C} - \boldsymbol{x}_0 &= 0 \, , \\
\frac{d}{dt} \boldsymbol{\phi}(t_l) \boldsymbol{C} - \boldsymbol{f}(\boldsymbol{\phi}(t_l) \boldsymbol{C},u(t_l),\boldsymbol{\vartheta}) &= 0 \,, \\
\bigtriangledown \boldsymbol{\Lambda(C,\vartheta,\lambda)} &= 0 \, .
\end{align}
\noindent {\bf Trajectory Matching Methods} \hfill \\

\textit{Trajectory Matching} methods are the well-known least squares formulation, which is equivalent to Equation (\ref{eq:trajectorySSE}). Fitting data with SSE is reasonable; however, there are some drawbacks:
\begin{itemize}
\item \textbf{Multiple Local Minima on a Complex Surface: }It is very likely that fitting the data requires the optimization of the parameters over a complex space with multiple local minima, and thus, the results tend to get trapped instead of moving toward the global minimum.
\item \textbf{Costly Computation:} Repeatedly solving ODEs at different parameter values and initial conditions can be computationally costly, and it is often a problem seen in highly-dimensional nonlinear dynamics. 
\end{itemize}

\medskip

\noindent {\bf Gradient Matching Methods} \\

\textit{Gradient matching} method is available to solve ODEs in contrast to \textit{trajectory matching}. In gradient matching method, instead of solving the differential equations through data-fitting, gradient matching method minimizes the derivatives' errors: 
\begin{equation}
ISSE(\boldsymbol{\vartheta}) = \int \parallel \frac{d}{dt} \boldsymbol{x}(t) - \boldsymbol{f(x,u,\vartheta)} \parallel^2 \, dt.
\end{equation}
Gradient matching avoids the problems that trajectory matching encounters from three aspects: less costly computation, more accurate results, and less bias. Nevertheless, the problem in gradient matching is the requirement of sufficiently many data to estimate both $\boldsymbol{x}$ and $\frac{d}{dt} \boldsymbol{x}$. \\

\noindent \textbf{Smoothing Methods and Basis Expansions} \\

\textit{Smoothing methods} suggest that not only the goodness of data-fitting but also the measure of the complexity should be plugged into the objective function; therefore, we arrive at the tweaked objective function:
\begin{equation}
SSSE(\boldsymbol{c}) = (1-\rho) \sum_{j=1}^{n} (g^{-1}(y_j) - \phi(t_j)^T \boldsymbol{c} )^2 + \rho \int \big[L \boldsymbol{\phi}(t)^T \boldsymbol{c}]^2 \, dt, \quad 0 < \rho < 1 \ .
\end{equation}
The value of $\rho$ determines the trade-off between the goodness of fit of data and the complexity. Subsequently, the solution of the coefficient vector $\boldsymbol{c}$ is reformulated as
\begin{equation}
\boldsymbol{\hat{c}} = ((1-\rho)\boldsymbol{\Phi^T \Phi} + \rho \boldsymbol{R} )^{-1} \boldsymbol{\Phi^T y} \, ,
\end{equation}
where $\boldsymbol{R}$ is the matrix with entries
\begin{equation}
\label{eq:R}
R_{kl} = \int L \phi_k (t)  L \phi_l (t) \, dt
\end{equation} 
and $L$ is the differential operator such that 
\begin{equation}
L [\boldsymbol{x}] = 0
\end{equation}

Plugging $\boldsymbol{\hat{c}}$ into the equation, we can estimate $\hat{x}(t)$ and $\frac{d}{dt}\hat{x}(t)$ as:
\begin{align}
\hat{x}(t) &= \boldsymbol{\phi}(t)^T \boldsymbol{\hat{c}}, \\
\frac{d}{dt}\hat{x}(t) &= \frac{d}{dt} \boldsymbol{\phi}(t)^T \boldsymbol{\hat{c}}.
\end{align}


\noindent {\bf Profiled Estimation} \hfill \\

\textit{Profiled Estimation} combines the spirits of trajectory matching and gradient matching methods together so as to avoid the disadvantages of both approaches; such technique utilizes \textit{parameter cascading} [\ref{alg:PC}], which is a tactic similar to expectation maximization method \cite{ramsay2017dynamic}. In each iteration, coefficient vector $\boldsymbol{c(\theta)}$ are optimized in the inner fitting criterion $\boldsymbol{\mathcal{J}(c|\theta)} $, and parameters $\boldsymbol{\theta}$ are optimized in the outer criterion $\mathcal{H} (\boldsymbol{\theta} )$.

In forcing systems, external inputs are considered, so the product of differential operator $L$ changes to
\begin{equation}
Lx(t) = \sum_{l}^{L^*} \alpha_{l | \boldsymbol{\theta} } (t) u_l (t) \, .
\end{equation}
Therefore, besides the matrix $\boldsymbol{R(\theta)}$ defined in Equation (\ref{eq:R}), a $K \times L^*$ matrix $\boldsymbol{S (\theta) }$ for inputs is defined as
\begin{equation}
\boldsymbol{S(\theta)} = \int_{0}^{T} [L \boldsymbol{\phi}(t) ] \boldsymbol{u}^T(t)  \, dt \, .
\end{equation}
The inner criterion $\mathcal{J}$ then becomes
\begin{equation}
\mathcal{J}(c | \theta) = (1 - \rho ) (\boldsymbol{g^{-1}(y)-\Phi c})^T (\boldsymbol{g^{-1}(y)-\Phi c}) /n + \rho \boldsymbol{c^T R (\theta) c} /T + \rho \boldsymbol{c^T S (\theta)} /T
\end{equation}
and the solution of the coefficient vector $c(\theta)$ is
\begin{equation}
\boldsymbol{c(\theta)} = [ (1-\rho) \boldsymbol{\Phi ^T \Phi}/n + \rho \boldsymbol{R(\theta) /T} ]^{-1} [ (1- \rho) \boldsymbol{\Phi^T g^{-1}(y)} /n + \rho \boldsymbol{S(\theta)}/T ]
\end{equation}
As for the outer optimization criterion $\mathcal{H}$,
\begin{equation}
\mathcal{H}(\boldsymbol{\theta} | \rho ) = G( \boldsymbol{g^{-1}(y),x(t)|\theta},\rho ) \, ,
\end{equation}
and function $G$ is user-dependent. 

To estimate the parameters $\boldsymbol{\theta}$, we seek $\boldsymbol{\theta}$ that minimize $\mathcal{H}(\boldsymbol{\theta}|\rho)$ and $c(\boldsymbol{\theta})$ that minimize $\mathcal{J}$ in each iteration. For outer criterion $\mathcal{H}(\boldsymbol{\theta}|\rho)$, we aim at solving 
\begin{equation}
\frac{d \mathcal{H}}{d \boldsymbol{\theta}}\big|_{\boldsymbol{\theta}} = 0 
\end{equation}
by using the implicit differentiation:
\begin{equation}
\frac{d \mathcal{H}}{d \boldsymbol{\theta}} = \frac{\partial \mathcal{H}}{\partial \boldsymbol{\theta}} + \frac{\partial \mathcal{H}}{\partial \boldsymbol{c}} \frac{d \boldsymbol{c}}{d \boldsymbol{\theta}} \, .
\end{equation}
In nonlinear system, it is often unable to express $\frac{d \boldsymbol{c}}{d \boldsymbol{\theta}}$ explicitly. However, suppose $\boldsymbol{c} $ is optimized to a certain degree, we are allowed to assume $\frac{\partial \mathcal{J} }{\partial \boldsymbol{c} } = 0$. Plugging it into the total derivative, we arrive at
\begin{equation}
\frac{d}{d \boldsymbol{\theta} } \bigg( \frac{\partial \mathcal{J} }{ \partial \boldsymbol{c} } \bigg) =  \frac{\partial^2 \mathcal{J}}{\partial \boldsymbol{c}\partial \boldsymbol{\theta} } + \frac{\partial^2 \mathcal{J}}{\partial \boldsymbol{c}^2} \frac{d \boldsymbol{c}}{d \boldsymbol{\theta}}.
\end{equation}
Then, $\frac{d \boldsymbol{c}}{d \boldsymbol{\theta}} $ can be expressed as 
\begin{equation}
\frac{d \boldsymbol{c}}{d \boldsymbol{\theta}} = - \Big( \frac{\partial^2 \mathcal{J}}{\partial \boldsymbol{c}^2} \Big)^{-1} \Big( \frac{\partial^2 \mathcal{J}}{\partial \boldsymbol{c}\partial \boldsymbol{\theta} } \Big) \, .
\end{equation}

For both optimization criteria $\mathcal{J}$ and $\mathcal{H}$, we can use profiled estimation to efficiently optimize $\boldsymbol{c}$ and $\boldsymbol{\theta}$ respectively.

\begin{algorithm}[H]
\caption{Parameter Cascading Algorithm}
\label{alg:PC}
\SetAlgoNoLine
Input: Criteria $\mathcal{J}$, $\mathcal{H}$, parameters $\boldsymbol{\theta}^0$\\
Output: Parameters $\boldsymbol{\theta}$\\
$N\_epoch=$ iteration upper bound \\
Initialize $\boldsymbol{\theta} = \boldsymbol{\theta}^0$
\While{$k<N\_epoch$}{
\Repeat{convergence}{
$\boldsymbol{c} = \argmin_{\boldsymbol{c}} \mathcal{J(\boldsymbol{c | \theta})} $ via Gauss-Newton Algorithm (\ref{alg:GN}) \\
$\boldsymbol{\theta} = \argmin_{\boldsymbol{\theta}} \mathcal{H(\boldsymbol{\theta | \rho})} $ via Gauss-Newton Algorithm (\ref{alg:GN}) \\
}
}
\end{algorithm}

\medskip

\noindent \textbf{From optimization's point of view:} The global optimum can always be found if the whole system is convex. Moreover, the problems can be made more approachable by reformulating the objective function with a tweak. The drawback is that the solution found may be a local optimum, and the initialization can also play an essential role in the process of some optimization methods.

\subsection{Encapsulation of Gauss-Newton, MCMC, and Variational Bayesian}
\label{sec:nmcritical}

We summarize the advantages and disadvantages of the three typical numerical methods, MCMC, VB, and Gauss-Newton with collocation methods in Table \ref{tab:nm}. Bayesian approaches include MCMC and Variational Bayesian (VB). Both contain the merits that beliefs are updated probabilistically, which makes the calculated state space reasonable and intuitive in each iteration. As for the disadvantages they bear, both are subjective to priors and the chosen objective function. In addition, MCMC takes long time to reach the stationary distribution or easily gets trapped at local minima. For VB, the derivation is often complicated and also prone to getting trapped at local minima. On the other hand, the Gauss-Newton algorithm together with collocation methods takes on the advantages such as high reliability, high feasibility, and a nearly convex surface in its objective function. However, such technique encounters the difficulties in making choices of the appropriate basis functions and the smoothing parameter.

\begin{table}[!htbp]
\begin{center}
\begin{tabularx}{\textwidth}{ |X|X|X|X| }
  \hline
  Methods & Gauss-Newton with Collocation Method & MCMC & Variational Bayesian (used in DCM) \\
  \hline 
  Advantages  & \begin{flushleft}
  \begin{itemize}
  \item High reliability
  \item Reasonably fast convergence
  \item Approximately convex surface in the objective function
  \item Feasible derivative calculation with basis functions
  \end{itemize}
  \end{flushleft} & \begin{flushleft}
  \begin{itemize}
  \item Beliefs can be updated in each iteration 
  \item User-defined probability
  \item Make use of conditional probability
  \end{itemize}
  \end{flushleft}  & \begin{flushleft}
  \begin{itemize}
  \item Beliefs can be updated in each iteration 
  \item User-defined probability
  \item Fast convergence for small datasets
  \end{itemize}
  \end{flushleft} \\
  \hline
  Disadvantages & \begin{flushleft}
  \begin{itemize}
  \item The choice of basis function matters
  \item The scale of the smoothing parameter has to be fine-tuned
  \end{itemize}
  \end{flushleft}  & \begin{flushleft}
  \begin{itemize}
  \item Subjective to priors
  \item Easily get trapped at local minima 
  \item Frequently take long time to reach the stationary distribution
  \end{itemize}
  \end{flushleft}  & \begin{flushleft}
  \begin{itemize}
  \item Subjective to the chosen objective function
  \item Complicated derivation 
  \item The global optimum is not guaranteed
  \end{itemize}
  \end{flushleft}  \\
  \hline
\end{tabularx}
\end{center}
\caption{Table of Dynamic Parameter Estimation Methods' Advantages / Disadvantages}
\label{tab:nm}
\end{table}
\FloatBarrier

\section{Evaluation of the Estimated Parameters} \label{sec:eval}

In this section, we survey the evaluation of the estimated parameters: going through the computation of $x(t)$ in the differential equations with the estimated parameters and discussing the method to make inference for parameters. \\
\begin{center}
{\bf Differential Equations and Systems}\\
\end{center}

Starting from linear differential equations with forcing inputs,
\begin{equation}
D \boldsymbol{x} = \boldsymbol{Ax} + \boldsymbol{u},
\end{equation}
the corresponding solution for the system becomes
\begin{equation}
\boldsymbol{x} (t) = \boldsymbol{x}(t_0) e^{t \boldsymbol{A}} + \int_{t_0}^{t} e^{(t-\tau) \boldsymbol{A} } \boldsymbol{u(\tau)} \, d \tau,
\end{equation}
where $x(t_0)$ is the initial condition.\\ Taking non-stationary systems into consideration, we have
\begin{equation}
D \boldsymbol{x} = \boldsymbol{A} (t) \boldsymbol{x},
\end{equation}
and its solution is
\begin{equation}
\boldsymbol{x} (t) = \boldsymbol{x}(t_0) \, exp \, [ - \int_{t_0}^{t} \boldsymbol{A(\tau)} \, d \tau ].
\end{equation}

We demonstrate the strategy of solving a nonlinear differential system piecewise linearly. Given the nonlinear differential system,
\begin{equation}
D \boldsymbol{x} = \boldsymbol{Ax} + \boldsymbol{u} + N(\boldsymbol{x}),
\end{equation}
where $N(\boldsymbol{x} )$ is the nonlinear component of the equation, we rewrite the formulation as a piecewise linear function:
\begin{align}
D \boldsymbol{x} &= \boldsymbol{A_0 x} + \boldsymbol{u_0} \, , \quad 0 \leq t < t_1 \\
D \boldsymbol{x} &= \boldsymbol{A_1 x} + \boldsymbol{u_1} \, , \quad t_1 \leq t < t_2 \\
& \qquad \qquad \vdots \\
D \boldsymbol{x} &= \boldsymbol{A_{N-1} x} + \boldsymbol{u_{N-1}} \, , \quad t_{N-1} \leq t < t_N \\
\end{align}
Following the piecewise linear functions, we are allowed to make use of the tactics in linear differential equations/systems for each linearized equation.

\begin{center}
{\bf Making Inference for Parameters} \\
\end{center}

\medskip
From the perspective of Bayesian statistics, the probabilities can infer how likely the event will occur; conversely, from the viewpoint of frequentists, the parameter set is presumed as a Gaussian distribution based on the central limit theorem,
\begin{equation}
\hat{\boldsymbol{\vartheta}} \sim \mathbb{N} \Bigg(\boldsymbol{\vartheta},\sigma^2 \big[\boldsymbol{J(\vartheta)^T J(\vartheta)} \big]^{-1} \Bigg),
\end{equation}
where $\hat{\boldsymbol{\vartheta}}$ is the estimated parameter set, and $\boldsymbol{\vartheta}$ the real parameter set; $\sigma^2$ is the population variance of the errors $\epsilon$. The estimation of $\sigma^2$ are given as the empirical residuals shown below:
\begin{equation}
\hat{\sigma}^2 = \frac{1}{n-p}\sum_{j=1}^{n}(g^{-1}(y_j)-x(t_j,\hat{\boldsymbol{\vartheta}}))^2,
\end{equation}
where $p$ is the number of estimated parameters and $n$ the number of data points. Based on the presumption that the error of $\hat{\boldsymbol{\vartheta}}-\boldsymbol{\vartheta}$ also follows a normal distribution,
\begin{equation}
\hat{\boldsymbol{\vartheta}}-\boldsymbol{\vartheta} \approx \mathbb{N} \Bigg(0,\sigma^2 \big[\boldsymbol{J(\vartheta)^T J(\vartheta)} \big]^{-1} \Bigg) \, .
\end{equation}
Following the logic, for example, if we want to know a particular parameter's estimated precision within $95 \%$ confidence interval, it could then be expressed as 
\begin{equation}
\big[ \hat{\vartheta}_k-1.96\hat{\sigma}_k \quad , \quad \hat{\vartheta}_k+1.96\hat{\sigma}_k \big] \, .
\end{equation}
To translate the results of estimated parameters into data prediction, we have the variance of $x(t;\boldsymbol{\vartheta})$
\begin{equation}
var(x(t;\boldsymbol{\vartheta})) \approx \boldsymbol{J(t;\boldsymbol{\vartheta})^T \sigma_{\vartheta}J(t;\boldsymbol{\vartheta})}
\end{equation}
Thus, to predict the measurement at time $t$ within $95 \%$ confidence interval, we arrive at
\begin{equation}
\big[ x(t;\hat{\vartheta})-1.96 (\sigma_x+\sigma) \quad , \quad x(t;\hat{\vartheta})+1.96 (\sigma_x+\sigma) \big] \, .
\end{equation}

\vspace{15mm}

\section{An Exemplary Problem Statement}
\label{sec:formulation}


In this section, we demonstrate how to formulate an effective network estimation problem: ERP is the selected dynamic model, and the interconnection among the cortical columns is based on the research conducted by Felleman et al. \cite{felleman1991distributed}. The goal is to search the parameters that best fit the assumed biophysical model with biologically reasonable constraints and given data (observations). First, the biological model of a basic neural population unit is described in \ref{npunit}. Next, the interconnection among these units is illustrated in \ref{interconnect}. Subsequently, the dynamic parameter estimation problem is stated in \ref{optimization}. Last, several biological features for evaluation are listed in \ref{sec:feature}.

\begin{center}
\subsection{ \textbf{Dynamic Model of the Human Brain} }
\label{npunit}
\end{center}


In our example, we still have Equations (1) and (2) represent neural transmission and experimental design, respectively. Function $f$ shows the mechanisms of ensemble neuronal transmission, whereas function $g$ represents the transformation from brain signals to the measured signals / data. Variables $x$ are the biological potentials in the brain; variables $u$ are the external stimuli along with the experiment design, which can be utilized to track the brain's transient response. Last, parameters $\theta$ indicate the strength, delay, speed, and interconnection strength of the neural signals.

\begin{figure}[!htpb]
  \centering
  \includegraphics[width=0.6\linewidth]{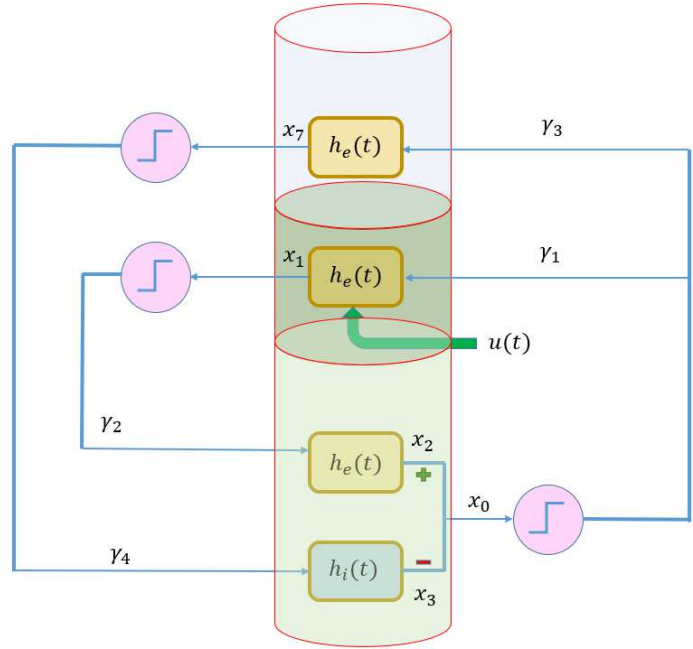}
  \caption{Schematic used to model single cortical column. Three cylinders stack together to represent supragranular layer, layer 4, and infragranular layer from top to bottom. $h_e (t)$ / $h_i (t)$ is the linear transformation with an impulse response of excitatory / inhibitory neuronal populations. $x$ are the average post-synaptic membrane potentials;$\gamma$ are the internal connection strengths; $u(t)$ is the external stimulus. Each pink circle denotes a sigmoid function, and the blue arrows show how signals transfer in the system}
  \label{fig:singlecolumn}
\end{figure}

The functioning architecture of a cortical column is explained in this paragraph along with our illustration (Figure \ref{fig:singlecolumn}). A cortical column is usually said to be the fundamental functional unit of the brain, and its simplified structure is demonstrated in \textbf{Figure \ref{fig:singlecolumn}}. \textbf{Figure \ref{fig:singlecolumn}} is created with \cite{jansen1995electroencephalogram,felleman1991distributed,david2006dynamic} as references. In \textbf{Figure \ref{fig:singlecolumn}}, a cortical column is represented by three stacked cylinders. The bottom cylinder is supragranular layer, and inhibitory interneurons such as GABAergic neurotransmitters are in this layer. The middle cylinder is layer 4, which is composed of spiny stellate cells; moreover, layer 4 is the layer that receives external stimulus. The top cylinder is infragranular layer, which has numerous pyramidal cells. $h_e (t)$ / $h_i (t)$ is the linear transformation with an impulse response of excitatory / inhibitory neuronal populations. $x_1 \, , x_2 \, , x_3 \, , x_7 $ are the average post-synaptic membrane potentials; $\gamma_1 \, , \gamma_2. \, , \gamma_3 \, , \gamma_4 $ are the internal connection strengths; $u(t)$ is the external stimulus. Each pink circle denotes a sigmoid function, and the blue arrows show the signal paths in the cortical column. 
The differential equations for the system portrayed in \textbf{Figure \ref{fig:singlecolumn}} can be written as:
\begin{align} \label{dcmstateeq}
\dot{x}_0(t) &= x_5(t) - x_6(t) \\
\dot{x}_1(t) &= x_4(t) \\
\dot{x}_2(t) &= x_5(t) \\
\dot{x}_3(t) &= x_6(t) \\
\dot{x}_4(t) &= \frac{H_e}{\tau_e}(\gamma_1 S(x_0(t))+ u(t)) - \frac{2x_4(t)}{\tau_e} -\frac{x_1(t)}{\tau_e^2} \\
\dot{x}_5(t) &= \frac{H_e}{\tau_e}(\gamma_2S(x_1(t))) - \frac{2x_5(t)}{\tau_e} -\frac{x_2(t)}{\tau_e^2} \\
\dot{x}_6(t) &= \frac{H_i}{\tau_i}\gamma_4S(x_7(t))-\frac{2x_6(t)}{\tau_i}-\frac{x_3(t)}{\tau_i^2} \\
\dot{x_7}(t) &= x_8(t) \\
\dot{x_8}(t) &= \frac{H_e}{\tau_e}(\gamma_3S(x_0(t))) - \frac{2x_8(t)}{\tau_e} -\frac{x_7(t)}{\tau_e^2}, 
\end{align}
in which function $S$ is the sigmoid function, $H_e \ , \tau_e \ , H_i, \ \tau_i, \ $ are the parameters for membrane potentials, and $\gamma_1 \, , \gamma_2 \, , \gamma_3 \, , \gamma_4 $ are the coefficients for internal connection. The derivation of differential equations (76) - (84) are based on the neural mass model (NMM) proposed by Jansen and Rit in 1995 \cite{jansen1995electroencephalogram}. In NMM, the second order differential equation 
\begin{equation}
\ddot{x}(t)=Aau(t)-2a\dot{x}(t)-a^2x(t)
\end{equation} transforms average density of presynaptic inputs to postsynaptic membrane potential (PSP), where $u(t)$ and $x(t)$ are the presynaptic input and postsynaptic output signals, respectively; $A$, $a$ are the amplitude and decay time respectively. In the nine equations, \{$x_1,x_4$\}, \{$x_2,x_5$\}, \{$x_3,x_6$\}, \{$x_7,x_8$\} are the four ensemble neuronal populations represented with second order differential equations of an identical structure. Therefore, for excitatory neuronal populations, \{$x_1,x_4$\}, \{$x_2,x_5$\}, \{$x_7,x_8$\}, the solution is
\begin{equation}
h_e(t) = 
  \begin{cases}
    \frac{H_e}{\tau_e} te^{- \frac{t}{\tau_e} } \, , \, t \geq 0 \\
     0 \, , \, t< 0 .
  \end{cases}
\end{equation}
On the contrary, the solution for inhibitory interneurons, \{$x_3,x_6$\}, is
\begin{equation}
h_i(t) = 
  \begin{cases}
    \frac{H_i}{\tau_i} te^{- \frac{t}{\tau_i} } \, , \, t \geq 0 \\
     0 \, , \, t< 0 
  \end{cases}
\end{equation}
To successfully fire the neuron and complete signal propagation, the net presynaptic inputs must be larger than the threshold. Such threshold-dependent mechanism is described with a modified sigmoid function: 
\begin{equation}
S(x) = \frac{1}{1+exp(-rx)}-\frac{1}{2},
\end{equation}
where $r=0.56$. The internal connection strengths are set as $\gamma_1=C \, , \gamma_2=1.25C \, , \gamma_3=0.25C \, , \gamma_4=0.25C $, where constant $C$ depends on different functioning modes of the brain. For example, $C=135$ is for alpha generation \cite{jansen1995electroencephalogram}.\\

\subsection{ \textbf{Interconnection of Neuronal Populations} }
\label{interconnect}

\begin{figure}[!htpb]
  \centering
    \includegraphics[width=1.0\linewidth]{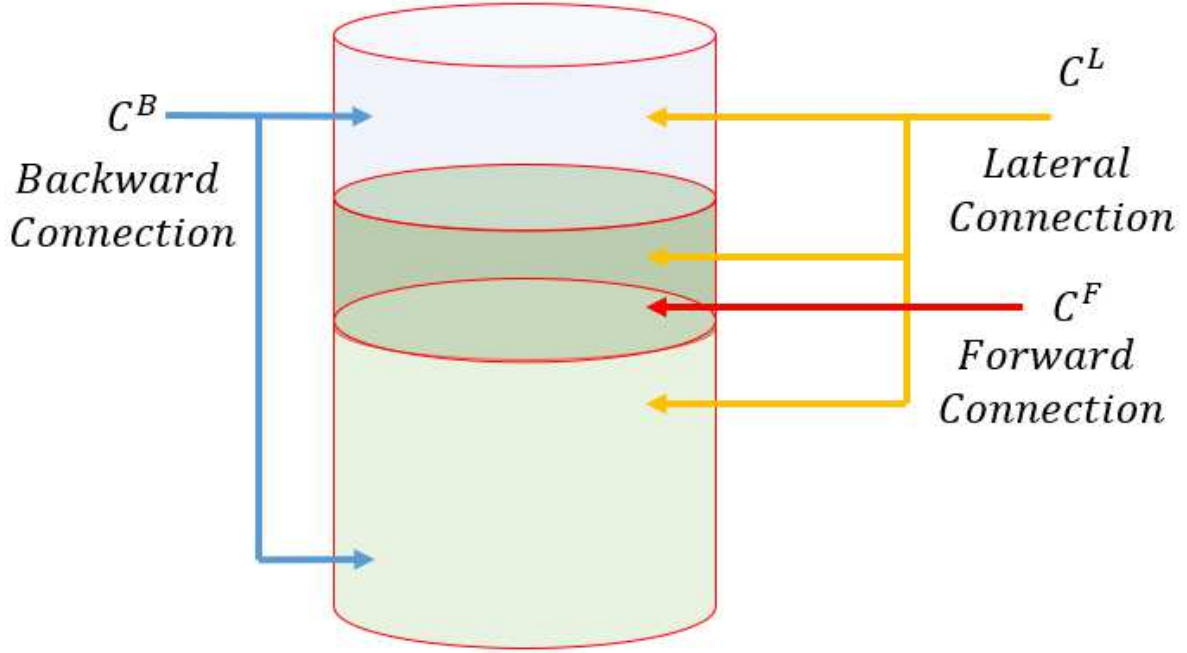}
  \caption{Schematic used to model cortical connectivity} 
  \label{fig:cortex_hierarchy}
\end{figure}

Dynamic brain mechanisms result from network interactions of multiple cortices, and effective interconnection consists of both intrinsic and extrinsic connections. Intrinsic connections ($\gamma_1$, $\gamma_2$, $\gamma_3$, $\gamma_4$) are fine-tuned by alpha activity \cite{jansen1995electroencephalogram}. Extrinsic connections, on the other hand, rest on a tri-partitioning of the cortical sheet into supragranular/intragranular layers and granular layer 4, which have been derived from visual cortex of macaques (Fig. \ref{fig:cortex_hierarchy}) \cite{felleman1991distributed}. The generalization of visual cortex is used in this study to represent other cortices. Under such simplified assumption, \textbf{Figure \ref{fig:cortex_hierarchy}} displays the interconnection among multiple cortical columns: forward, lateral, and backward connections. The arrangement of the stacked cylinders in \textbf{Figure \ref{fig:cortex_hierarchy}} is the same as in \textbf{Figure \ref{fig:singlecolumn} }; from top to bottom, they are infragranular layer, granular layer 4, and supragranular layer. Three directed interconnections include: 
\begin{enumerate}
\item Forward Connections ($C^F$, red): originate in agranular (both supra- and infra- granular) layers and terminate in layer 4;
\item Backward Connections ($C^B$, blue): originate in agranular layers; terminate in supragranular and infragranular layers;
\item Lateral Connections ($C^L$, yellow): originate in agranular layers and terminate in all layers.
\end{enumerate}

Incorporating interconnection into the single cortical column model, we can rewrite the differential equations as:
\begin{align}
\dot{x}_0(t) &= x_5(t) - x_6(t) \\
\dot{x}_1(t) &= x_4(t) \\
\dot{x}_2(t) &= x_5(t) \\
\dot{x}_3(t) &= x_6(t) \\
\dot{x}_4(t) &= \frac{H_e}{\tau_e}((C^F)(x_0(t-d_F))+(C^L)(x_0(t-d_L))+(\gamma_1I)S(x_0(t))+C^u u(t-d_u )) - \frac{2x_4(t)}{\tau_e} -\frac{x_1(t)}{\tau_e^2} \\
\dot{x}_5(t) &= \frac{H_e}{\tau_e}((C^B)(x_0(t-d_B))+(C^L)(x_0(t-d_L))+(\gamma_2 I )S(x_1(t))) - \frac{2x_5(t)}{\tau_e} -\frac{x_2(t)}{\tau_e^2} \\
\dot{x}_6(t) &= \frac{H_i}{\tau_i} (\gamma_4 I)S(x_7(t))-\frac{2x_6(t)}{\tau_i}-\frac{x_3(t)}{\tau_i^2} \\
\dot{x_7}(t) &= x_8(t) \\
\dot{x_8}(t) &= \frac{H_e}{\tau_e}((C^B) (x_0(t-d_B))+(C^L) (x_0(t-d_L))+(\gamma_3I)S(x_0(t))) - \frac{2x_8(t)}{\tau_e} -\frac{x_7(t)}{\tau_e^2} 
\end{align}
Equations (89)-(97) are the dynamic system with interconnection taken into account. Different from the single cortical column, variables $x$ become column vectors with each row symbolizing the neuronal potential/current of one cortical column. Matrices $C^F \ , C^L \ , C^B $ represent the forward, lateral, and backward connectivity strengths and $d_F \ , d_L \ , d_B $ are their corresponding distal delays. Connection matrix $C^u$, vector $u$, and stimuli delay $d_u$ are the characteristics of external inputs.




\subsection{ \textbf{Optimization Problem} }
\label{optimization}

We formulate the parameter estimation problem into an optimization problem 
\begin{equation}
\argmin_{\theta} \, (1- \lambda) (Y-g(x,\theta))^T (Y-g(x,\theta)) + \lambda (X - f(x,u,\theta) )^T (X - f(x,u,\theta) ) 
\end{equation}
subject to \\
\begin{flalign*}
    f&:neural \ mass \ model \\
    g&:observation \ function \\
    u&: designed \ inputs \\
    Y&: observations \ (neuroimaging \ data), Y=g(X,\theta) \\
    X&: potentials/currents \ of \ neuronal \ populations, \ a \ subset \ of \ X \ is \ known 
\end{flalign*}

$\theta = \{ H_e, \  H_i, \  C^F \ , C^L \ , C^B, \ C^u, \ d_F \ , d_L \ , d_B, \ d_u \}$ \\
$\tau_e = 10 (ms), \ \tau_i = 20 (ms), \ \gamma_1=C , \ \gamma_2=1.25C, \ \gamma_3=0.25C, \ \gamma_4=0.25C, \ C=135$ \\
$1 \geq \lambda \geq 0$ \\
Details of the givens and estimated parameters are listed in \textbf{Table \ref{tab:eqsort}}, including the summarized description of each variable's biological representation. Intrinsic and extrinsic propagation strengths and external delays are the parameters to be estimated. We assume the intrinsic connection and internal delay to be constants as defined in NMM. Additionally, state variables $x$ and observations (neuroimaging data) $y$ are capitalized in Equation (98) to be identified as givens.

\begin{table}[!htbp]
\newcolumntype{b}{X}
\newcolumntype{s}{>{\hsize=.5\hsize}X}
\newcolumntype{k}{>{\hsize=.25\hsize}X}
\begin{center}
\begin{tabularx}{\textwidth}{ kssb }
  \hline
  Variable(s) & Given / Target  & Location & Description \\
  \hline 
  $x_0$  & 
  Given
   & Supragranular layer
   & \begin{flushleft}
  \begin{itemize}
  \item The output signal of a single node
  \item $g(x_0,\theta)$ is the measured data
  \end{itemize}
  \end{flushleft} \\
  \hline
  $x_1 \, , \, x_4$ & Possibly Infeasible  & Spiny stellate cells at granular layer 4  & \begin{flushleft}
  \begin{itemize}
  \item $x_1$, $x_4$ together form a second order forcing system 
  \item Excitatory neuronal population 
  \item Receiving intrinsic signal $x_0$ and extrinsic signal $u$
  \end{itemize}
  \end{flushleft}  \\
  \hline
  $x_2 \, , \, x_5$ & Possibly Infeasible  & Excitatory interneurons at supragranular layer  & \begin{flushleft}
  \begin{itemize}
  \item $x_2$, $x_5$ together form a second order forcing system 
  \item Excitatory neuronal population 
  \item Receiving intrinsic signal $x_1$ and extrinsic signal $x_0$
  \end{itemize}
  \end{flushleft}  \\
  \hline
  $x_3 \, , \, x_6$ & Possibly Infeasible  & Inhibitory interneurons at supragranular layer  & \begin{flushleft}
  \begin{itemize}
  \item $x_3$, $x_6$ together form a second order forcing system 
  \item Inhibitory neuronal population 
  \item Receiving intrinsic signal $x_7$
  \end{itemize}
  \end{flushleft}  \\
  \hline
  $x_7 \, , \, x_8$ & Possibly Infeasible  & Pyramidal cells at infragranular layer  & \begin{flushleft}
  \begin{itemize}
  \item $x_7$, $x_8$ together form a second order forcing system 
  \item Excitatory neuronal population 
  \item Receiving both intrinsic and extrinsic signal $x_0$
  \end{itemize}
  \end{flushleft}  \\
  \hline
  $C^F, \, C^B, \, C^L$ & Target & & Interconnection between cortical columns \\
  \hline
  $C^u, \, d_u$ & Target & & Parameters of external inputs \\
  \hline
  $u$ & Given (lab design) & & External inputs / stimulus \\
  \hline
  $H_e, \, H_i$ & Target, Priors Given (NMM) & & Magnitude of excitatory / inhibitory signal  \\
  \hline
  $\tau_e, \, \tau_i$ & Given (NMM) & & decay time of excitatory / Inhibitory signal  \\
  \hline
  $\gamma_1, \, \gamma_2, \, \gamma_3, \, \gamma_4$ &  Given (NMM) & & Intrinsic connection strength \\
  \hline
  $d_F, \, d_L, \, d_B$ & Target & & Delays between columns / regions \\
  \hline
\end{tabularx}
\end{center}
\caption{Table of Descriptions of Parameters in Exemplary Problem Statement}
\label{tab:eqsort}
\end{table}
\FloatBarrier

\subsection{ Features for Evaluation}
\label{sec:feature}
In our application, features related to effective connections under certain function/mechanism of a human brain are of our interest, and we show how these features can be obtained through the estimated parameters in the following items. In addition, we include the approaches made in \textit{DCM}, which is currently the most popular package developed by Karl Friston and associates \cite{david2006dynamic}. Features of our interest include:

\begin{itemize}
\item \textbf{Causality:} Parameters $C^F, \ C^L, \ C^B$ enable the estimation of causality in the brain network; their magnitude implies the interconnection among the selected cortical columns. In \textit{DCM}, users have to first specify the hypothesized connection among the chosen regions from the data, and \textit{DCM} will return the probability of the assumed connectivity.
\item \textbf{Propagation Delay:} Parameters $d_F, \ d_L, \ d_B, \ d_u$ stand for the propagation delays in the network interaction. In \textit{DCM}, propagation delays are assumed to be sufficiently small such that Taylor expansion around the specified time point is still precise.
\item \textbf{Diverse Neuronal Type:} Parameters $H_e, \ H_i$ represent the amplitudes of different neuronal population, and identical assumption is utilized in \textit{DCM}.
\item \textbf{Biologically Acceptable Estimation:} The decay time $\tau_e, \ \tau_i$ and internal connection strengths $\gamma_1, \gamma_2, \gamma_3, \gamma_4$ are fixed so as to meet the criteria of neurology. In \textit{DCM}, all of these parameters are sampled from a biologically reasonable sampling space.
\item \textbf{Stability of the brain:} Stability can infer the functioning wellness of the human brain. By making use of the collocation method, we can estimate the stability of the whole system through the choice of the basis functions and the evaluation of their corresponding coefficient vectors.
\end{itemize}

\section{Future Work}
\label{sec:future}

Currently, the most popular dynamic parameter estimation model of the brain, \textit{DCM}, leaves some space for improvement, and we propose to use the constrained optimization methods to construct a more robust and flexible dynamic model for the estimation of the brain mechanisms. 


\begin{enumerate}
\item \textbf{Allow Non-stationarity: } \textit{DCM} assumes the system being stationary in a given brain activity. However, in a particular activity, the brain experiences different states, and the assumption of non-stationarity produces better results than stationary one as evidenced in \cite{kim2017state}. In contrast to \textit{DCM}'s assumption, a non-stationary dynamic system of a brain is more realistic. With the collocation method, we can assume the parameters as a function of time or other parameters.
\begin{align}
\{H_e,\, H_i \} & \rightarrow \{H_e(t),\, H_i(t) \}
\end{align}
\item \textbf{Handle Stability Issue: } Stability is not tackled in \textit{DCM}, and one of the reasons is that it only applies trajectory matching to its model's optimization; furthermore, the brain is a nonlinear system, yet Friston et al.'s use linear approximation under the assumption that the neuronal activity is weakly nonlinear. In \cite{friston2002bayesian}, the stability issue is addressed, and cannot be handled if the given priors are incorrect. In contrast to the complicated functional analysis in \textit{DCM}, we have a simpler and more flexible approach to deal with the stability issue.
\item \textbf{Delay Estimation: }Delay issue is tackled in Friston et al.'s works, but the approach is deficient. It assumes the delay is small enough to take a Taylor expansion:
\begin{align}
\dot{x}_{i}(t) &= f_i(x_1(t-\tau_{i1}),...,x_n(t-\tau_{in}) ) \\
\dot{x}_{i}(t) &= f_i(x(t))-\sum_{j} \tau_{ij}\frac{\partial f_i}{\partial \tau_{ij}} \\
&= f_i(x(t))-\sum_{j} \tau_{ij}J_{ij}\dot{x}(t)_{j}
\end{align}
Nonetheless, the delay is often large enough to refute the hypothesis in real cases.
\textit{DCM} assumes the delay is sufficiently small such that Taylor expansion around the specified time point is still precise. Standing on the fact that delay is not differentiable but reflects in the data itself, it is feasible to tackle it in the data correlation terms through different strategies such as machine learning and neural network methods.

\item \textbf{Fix the Local Minima Problem: }Constrained optimization is chosen over VB in the proposed future work. Every optimization method has its merit and weakness. For VB used in \textit{DCM}, the finding of global optimum is not guaranteed since it is highly dependent of the initial point. Moreover, the derivation of the outcome is difficult. Last, VB applies functional analysis by using a novel distribution $q(\theta)$ to approximate the real distribution $p(\theta | y,m)$ based on the observables $y$. If the proposed distribution $q(\theta)$ cannot approximate the real distribution well, the results will be bad. Compared to VB, the constrained optimization can optimize over an approximately convex space given a good choice of the smoothing parameter. Moreover, with the simplicity and flexibility of constrained optimization, we can combine trajectory matching and gradient matching together in the objective function. 
\item \textbf{Include a Biologically Reasonable Brain Mapping: } With the growth in the research on diffusion tensor imaging (DTI), the connection structure in human brains gradually comes to light. Dynamic parameter estimation becomes more reliable with the structural brain mapping taken into consideration as the constraint of the optimization problem so as to fully utilize the time derivative information from the data.
\item \textbf{Welcome the era of eAI: }eAI is an acronym for emotional artificial intelligence. When the brain functions, a person's emotion also plays the role in the processing mechanisms \cite{minsky2007emotion}. For example, the selection of some parameters should be constrained not only physiologically but also psychologically. Therefore, We would like to incorporate emotions into our future work. 
\end{enumerate}

\section{Conclusions}
\label{sec:conclude}

State-of-the-art approaches for dynamic parameter estimation of brain mechanisms are covered in this report. We have surveyed the dynamic models, effective network estimation techniques, and dynamic parameter estimation methods. Moreover, we demonstrate the formulation of an effective network estimation problem and provide several future directions. Demystifying brain mechanisms has still been a tough challenge, and hopefully, the future work identified can resolve some of the puzzles.

\section{Acknowledgments}

The author would like to thank her research advisor, Chung-Kuan Cheng, for his support and feedback on the report. Also, thank you to my family and friends for making the report understandable. Last, I want to show my thankfulness to my research committee for taking the time to review my exam and broadening my horizon.


\bibliographystyle{unsrt}  
\bibliography{references}  

\begin{thebibliography}{10}

\bibitem{razi2016connected}
Adeel Razi and Karl~J Friston.
\newblock The connected brain: causality, models, and intrinsic dynamics.
\newblock {\em IEEE Signal Processing Magazine}, 33(3):14--35, 2016.

\bibitem{van2013wu}
David~C Van~Essen, Stephen~M Smith, Deanna~M Barch, Timothy~EJ Behrens, Essa
  Yacoub, Kamil Ugurbil, Wu-Minn~HCP Consortium, et~al.
\newblock The wu-minn human connectome project: an overview.
\newblock {\em Neuroimage}, 80:62--79, 2013.

\bibitem{insel2013nih}
Thomas~R Insel, Story~C Landis, and Francis~S Collins.
\newblock The nih brain initiative.
\newblock {\em Science}, 340(6133):687--688, 2013.

\bibitem{bressler2010large}
Steven~L Bressler and Vinod Menon.
\newblock Large-scale brain networks in cognition: emerging methods and
  principles.
\newblock {\em Trends in cognitive sciences}, 14(6):277--290, 2010.

\bibitem{sharkey2009connectionism}
Amanda~JC Sharkey and Noel Sharkey.
\newblock Connectionism.
\newblock In {\em The Routledge Companion to Philosophy of Psychology}, pages
  202--214. Routledge, 2009.

\bibitem{tononi1994measure}
Giulio Tononi, Olaf Sporns, and Gerald~M Edelman.
\newblock A measure for brain complexity: relating functional segregation and
  integration in the nervous system.
\newblock {\em Proceedings of the National Academy of Sciences},
  91(11):5033--5037, 1994.

\bibitem{cole2013multi}
Michael~W Cole, Jeremy~R Reynolds, Jonathan~D Power, Grega Repovs, Alan
  Anticevic, and Todd~S Braver.
\newblock Multi-task connectivity reveals flexible hubs for adaptive task
  control.
\newblock {\em Nature neuroscience}, 16(9):1348, 2013.

\bibitem{ketchum1959mind}
JD~Ketchum.
\newblock Mind and mechanism, 1959.
\newblock {\em The Canadian Psychologist}, 8(4):78, 1959.

\bibitem{hubel1974uniformity}
David~H Hubel and Torsten~N Wiesel.
\newblock Uniformity of monkey striate cortex: a parallel relationship between
  field size, scatter, and magnification factor.
\newblock {\em Journal of Comparative Neurology}, 158(3):295--305, 1974.

\bibitem{wiesel1974ordered}
Torsten~N Wiesel and David~H Hubel.
\newblock Ordered arrangement of orientation columns in monkeys lacking visual
  experience.
\newblock {\em Journal of comparative neurology}, 158(3):307--318, 1974.

\bibitem{wurtz2009recounting}
Robert~H Wurtz.
\newblock Recounting the impact of hubel and wiesel.
\newblock {\em The Journal of physiology}, 587(12):2817--2823, 2009.

\bibitem{george2009towards}
Dileep George and Jeff Hawkins.
\newblock Towards a mathematical theory of cortical micro-circuits.
\newblock {\em PLoS computational biology}, 5(10):e1000532, 2009.

\bibitem{lecun2015deep}
Yann LeCun, Yoshua Bengio, and Geoffrey Hinton.
\newblock Deep learning.
\newblock {\em nature}, 521(7553):436, 2015.

\bibitem{friston2002functional}
Karl Friston.
\newblock Functional integration and inference in the brain.
\newblock {\em Progress in neurobiology}, 68(2):113--143, 2002.

\bibitem{macaluso2005multisensory}
Emiliano Macaluso and Jon Driver.
\newblock Multisensory spatial interactions: a window onto functional
  integration in the human brain.
\newblock {\em Trends in neurosciences}, 28(5):264--271, 2005.

\bibitem{garrido2009mismatch}
Marta~I Garrido, James~M Kilner, Klaas~E Stephan, and Karl~J Friston.
\newblock The mismatch negativity: a review of underlying mechanisms.
\newblock {\em Clinical neurophysiology}, 120(3):453--463, 2009.

\bibitem{friston2011functional}
Karl~J Friston.
\newblock Functional and effective connectivity: a review.
\newblock {\em Brain connectivity}, 1(1):13--36, 2011.

\bibitem{greicius2003functional}
Michael~D Greicius, Ben Krasnow, Allan~L Reiss, and Vinod Menon.
\newblock Functional connectivity in the resting brain: a network analysis of
  the default mode hypothesis.
\newblock {\em Proceedings of the National Academy of Sciences},
  100(1):253--258, 2003.

\bibitem{wibral2014transfer}
Michael Wibral, Raul Vicente, and Michael Lindner.
\newblock Transfer entropy in neuroscience.
\newblock In {\em Directed information measures in neuroscience}, pages 3--36.
  Springer, 2014.

\bibitem{aertsen1991dynamics}
AHMJ Aertsen.
\newblock Dynamics of activity and connectivity in physiological neuronal
  networks.
\newblock {\em Nonlinear dynamics and neuronal networks}, 1991.

\bibitem{villaverde2014reverse}
Alejandro~F Villaverde and Julio~R Banga.
\newblock Reverse engineering and identification in systems biology:
  strategies, perspectives and challenges.
\newblock {\em Journal of the Royal Society Interface}, 11(91):20130505, 2014.

\bibitem{david2006dynamic}
Olivier David, Stefan~J Kiebel, Lee~M Harrison, J{\'e}r{\'e}mie Mattout,
  James~M Kilner, and Karl~J Friston.
\newblock Dynamic causal modeling of evoked responses in eeg and meg.
\newblock {\em NeuroImage}, 30(4):1255--1272, 2006.

\bibitem{baillet2001electromagnetic}
Sylvain Baillet, John~C Mosher, and Richard~M Leahy.
\newblock Electromagnetic brain mapping.
\newblock {\em IEEE Signal processing magazine}, 18(6):14--30, 2001.

\bibitem{moran2013neural}
Rosalyn~J Moran, Dimitris~A Pinotsis, and Karl~J Friston.
\newblock Neural masses and fields in dynamic causal modeling.
\newblock {\em Frontiers in computational neuroscience}, 7:57, 2013.

\bibitem{freeman1987simulation}
Walter~J Freeman.
\newblock Simulation of chaotic eeg patterns with a dynamic model of the
  olfactory system.
\newblock {\em Biological cybernetics}, 56(2-3):139--150, 1987.

\bibitem{jansen1995electroencephalogram}
Ben~H Jansen and Vincent~G Rit.
\newblock Electroencephalogram and visual evoked potential generation in a
  mathematical model of coupled cortical columns.
\newblock {\em Biological cybernetics}, 73(4):357--366, 1995.

\bibitem{wendling2000relevance}
Fabrice Wendling, Jean-Jacques Bellanger, Fabrice Bartolomei, and Patrick
  Chauvel.
\newblock Relevance of nonlinear lumped-parameter models in the analysis of
  depth-eeg epileptic signals.
\newblock {\em Biological cybernetics}, 83(4):367--378, 2000.

\bibitem{felleman1991distributed}
Daniel~J Felleman and DC~Essen Van.
\newblock Distributed hierarchical processing in the primate cerebral cortex.
\newblock {\em Cerebral cortex (New York, NY: 1991)}, 1(1):1--47, 1991.

\bibitem{garrido2008functional}
Marta~I Garrido, Karl~J Friston, Stefan~J Kiebel, Klaas~E Stephan, Torsten
  Baldeweg, and James~M Kilner.
\newblock The functional anatomy of the mmn: a dcm study of the roving
  paradigm.
\newblock {\em Neuroimage}, 42(2):936--944, 2008.

\bibitem{boly2011preserved}
Melanie Boly, Marta~Isabel Garrido, Olivia Gosseries, Marie-Aur{\'e}lie Bruno,
  Pierre Boveroux, Caroline Schnakers, Marcello Massimini, Vladimir Litvak,
  Steven Laureys, and Karl Friston.
\newblock Preserved feedforward but impaired top-down processes in the
  vegetative state.
\newblock {\em Science}, 332(6031):858--862, 2011.

\bibitem{naatanen2005memory}
Risto N{\"a}{\"a}t{\"a}nen, Thomas Jacobsen, and Istv{\'a}n Winkler.
\newblock Memory-based or afferent processes in mismatch negativity (mmn): A
  review of the evidence.
\newblock {\em Psychophysiology}, 42(1):25--32, 2005.

\bibitem{friston2012dcm}
Karl~J Friston, A~Bastos, Vladimir Litvak, Klaas~E Stephan, Pascal Fries, and
  Rosalyn~J Moran.
\newblock Dcm for complex-valued data: cross-spectra, coherence and
  phase-delays.
\newblock {\em Neuroimage}, 59(1):439--455, 2012.

\bibitem{whittington1995synchronized}
Miles~A Whittington, Roger~D Traub, and John~GR Jefferys.
\newblock Synchronized oscillations in interneuron networks driven by
  metabotropic glutamate receptor activation.
\newblock {\em Nature}, 373(6515):612, 1995.

\bibitem{moran2008bayesian}
Rosalyn~J Moran, Klaas~E Stephan, Stefan~J Kiebel, N~Rombach, William~T
  O'Connor, KJ~Murphy, RB~Reilly, and Karl~J Friston.
\newblock Bayesian estimation of synaptic physiology from the spectral
  responses of neural masses.
\newblock {\em Neuroimage}, 42(1):272--284, 2008.

\bibitem{hodgkin1952components}
Allan~L Hodgkin and Andrew~F Huxley.
\newblock The components of membrane conductance in the giant axon of loligo.
\newblock {\em The Journal of physiology}, 116(4):473--496, 1952.

\bibitem{morris1981voltage}
Catherine Morris and Harold Lecar.
\newblock Voltage oscillations in the barnacle giant muscle fiber.
\newblock {\em Biophysical journal}, 35(1):193--213, 1981.

\bibitem{fitzhugh1961impulses}
Richard FitzHugh.
\newblock Impulses and physiological states in theoretical models of nerve
  membrane.
\newblock {\em Biophysical journal}, 1(6):445--466, 1961.

\bibitem{nagumo1962active}
Jinichi Nagumo, Suguru Arimoto, and Shuji Yoshizawa.
\newblock An active pulse transmission line simulating nerve axon.
\newblock {\em Proceedings of the IRE}, 50(10):2061--2070, 1962.

\bibitem{jones2007neural}
Stephanie~R Jones, Dominique~L Pritchett, Steven~M Stufflebeam, Matti
  H{\"a}m{\"a}l{\"a}inen, and Christopher~I Moore.
\newblock Neural correlates of tactile detection: a combined
  magnetoencephalography and biophysically based computational modeling study.
\newblock {\em Journal of Neuroscience}, 27(40):10751--10764, 2007.

\bibitem{moran2011consistent}
Rosalyn~J Moran, Klaas~E Stephan, Raymond~J Dolan, and Karl~J Friston.
\newblock Consistent spectral predictors for dynamic causal models of
  steady-state responses.
\newblock {\em Neuroimage}, 55(4):1694--1708, 2011.

\bibitem{marreiros2010dynamic}
Andr{\'e}~C Marreiros, Stefan~J Kiebel, and Karl~J Friston.
\newblock A dynamic causal model study of neuronal population dynamics.
\newblock {\em Neuroimage}, 51(1):91--101, 2010.

\bibitem{moran2011vivo}
Rosalyn~J Moran, Mkael Symmonds, Klaas~E Stephan, Karl~J Friston, and Raymond~J
  Dolan.
\newblock An in vivo assay of synaptic function mediating human cognition.
\newblock {\em Current Biology}, 21(15):1320--1325, 2011.

\bibitem{pinotsis2012dynamic}
Dimitris~A Pinotsis, Rosalyn~J Moran, and Karl~J Friston.
\newblock Dynamic causal modeling with neural fields.
\newblock {\em Neuroimage}, 59(2):1261--1274, 2012.

\bibitem{pinotsis2013dynamic}
Dimitris~A Pinotsis, Dietrich~Samuel Schwarzkopf, Vladimir Litvak, Geraint
  Rees, G~Barnes, and Karl~J Friston.
\newblock Dynamic causal modelling of lateral interactions in the visual
  cortex.
\newblock {\em Neuroimage}, 66:563--576, 2013.

\bibitem{muthukumaraswamy2009resting}
Suresh~D Muthukumaraswamy, Richard~AE Edden, Derek~K Jones, Jennifer~B
  Swettenham, and Krish~D Singh.
\newblock Resting gaba concentration predicts peak gamma frequency and fmri
  amplitude in response to visual stimulation in humans.
\newblock {\em Proceedings of the National Academy of Sciences},
  106(20):8356--8361, 2009.

\bibitem{douglas1991functional}
Rodney~J Douglas and KA~Martin.
\newblock A functional microcircuit for cat visual cortex.
\newblock {\em The Journal of physiology}, 440(1):735--769, 1991.

\bibitem{packer2011dense}
Adam~M Packer and Rafael Yuste.
\newblock Dense, unspecific connectivity of neocortical parvalbumin-positive
  interneurons: a canonical microcircuit for inhibition?
\newblock {\em Journal of Neuroscience}, 31(37):13260--13271, 2011.

\bibitem{friston2003dynamic}
Karl~J Friston, Lee Harrison, and Will Penny.
\newblock Dynamic causal modelling.
\newblock {\em Neuroimage}, 19(4):1273--1302, 2003.

\bibitem{schuyler2010dynamic}
Brianna Schuyler, John~M Ollinger, Terrence~R Oakes, Tom Johnstone, and
  Richard~J Davidson.
\newblock Dynamic causal modeling applied to fmri data shows high reliability.
\newblock {\em Neuroimage}, 49(1):603--611, 2010.

\bibitem{friston2011network}
Karl~J Friston, Baojuan Li, Jean Daunizeau, and Klaas~E Stephan.
\newblock Network discovery with dcm.
\newblock {\em Neuroimage}, 56(3):1202--1221, 2011.

\bibitem{brown2012dynamic}
Harriet~R Brown and Karl~J Friston.
\newblock Dynamic causal modelling of precision and synaptic gain in visual
  perception—an eeg study.
\newblock {\em Neuroimage}, 63(1):223--231, 2012.

\bibitem{adams2016dynamic}
Rick~A Adams, Markus Bauer, Dimitris Pinotsis, and Karl~J Friston.
\newblock Dynamic causal modelling of eye movements during pursuit: confirming
  precision-encoding in v1 using meg.
\newblock {\em Neuroimage}, 132:175--189, 2016.

\bibitem{zhuang2005connectivity}
Jiancheng Zhuang, Stephen LaConte, Scott Peltier, Kan Zhang, and Xiaoping Hu.
\newblock Connectivity exploration with structural equation modeling: an fmri
  study of bimanual motor coordination.
\newblock {\em NeuroImage}, 25(2):462--470, 2005.

\bibitem{goebel2003investigating}
Rainer Goebel, Alard Roebroeck, Dae-Shik Kim, and Elia Formisano.
\newblock Investigating directed cortical interactions in time-resolved fmri
  data using vector autoregressive modeling and granger causality mapping.
\newblock {\em Magnetic resonance imaging}, 21(10):1251--1261, 2003.

\bibitem{dhamala2018granger}
Mukesh Dhamala, Hualou Liang, Steven~L Bressler, and Mingzhou Ding.
\newblock Granger-geweke causality: estimation and interpretation.
\newblock {\em NeuroImage}, 175:460--463, 2018.

\bibitem{sheikhattar2018extracting}
Alireza Sheikhattar, Sina Miran, Ji~Liu, Jonathan~B Fritz, Shihab~A Shamma,
  Patrick~O Kanold, and Behtash Babadi.
\newblock Extracting neuronal functional network dynamics via adaptive granger
  causality analysis.
\newblock {\em Proceedings of the National Academy of Sciences},
  115(17):E3869--E3878, 2018.

\bibitem{sivia2006data}
Devinderjit Sivia and John Skilling.
\newblock {\em Data analysis: a Bayesian tutorial}.
\newblock OUP Oxford, 2006.

\bibitem{tzikas2008variational}
Dimitris~G Tzikas, Aristidis~C Likas, and Nikolaos~P Galatsanos.
\newblock The variational approximation for bayesian inference.
\newblock {\em IEEE Signal Processing Magazine}, 25(6):131--146, 2008.

\bibitem{brown2004statistical}
Kevin~S Brown, Colin~C Hill, Guillermo~A Calero, Christopher~R Myers, Kelvin~H
  Lee, James~P Sethna, and Richard~A Cerione.
\newblock The statistical mechanics of complex signaling networks: nerve growth
  factor signaling.
\newblock {\em Physical biology}, 1(3):184, 2004.

\bibitem{gutenkunst2007universally}
Ryan~N Gutenkunst, Joshua~J Waterfall, Fergal~P Casey, Kevin~S Brown,
  Christopher~R Myers, and James~P Sethna.
\newblock Universally sloppy parameter sensitivities in systems biology models.
\newblock {\em PLoS computational biology}, 3(10):e189, 2007.

\bibitem{boyd2004convex}
Stephen Boyd and Lieven Vandenberghe.
\newblock {\em Convex optimization}.
\newblock Cambridge university press, 2004.

\bibitem{ramsay2017dynamic}
James Ramsay and Giles Hooker.
\newblock Dynamic data analysis, 2017.

\bibitem{kim2017state}
Seong-Eun Kim, Michael~K Behr, Demba Ba, and Emery~N Brown.
\newblock State-space multitaper time-frequency analysis.
\newblock {\em Proceedings of the National Academy of Sciences}, page
  201702877, 2017.

\bibitem{friston2002bayesian}
Karl~J Friston.
\newblock Bayesian estimation of dynamical systems: an application to fmri.
\newblock {\em NeuroImage}, 16(2):513--530, 2002.

\bibitem{minsky2007emotion}
Marvin Minsky.
\newblock {\em The emotion machine: Commonsense thinking, artificial
  intelligence, and the future of the human mind}.
\newblock Simon and Schuster, 2007.

\end{thebibliography}

\end{document}